\documentclass[useAMS,usenatbib]{mn2e}

\usepackage{graphicx}
\usepackage{longtable}
\usepackage{lscape}

\usepackage{natbib}     % for bibliography
 
\begin{document}

\title[Kinematic Structure in the Galactic Halo at the North Galactic Pole.]
{Kinematic Structure in the Galactic Halo at the North Galactic Pole:
 RR Lyrae and BHB Stars show different kinematics.}

\author[T.D. Kinman, C. Cacciari, A. Bragaglia, A. Buzzoni, A. Spagna]
{T.D. Kinman$^{1}$\thanks{e-mail: kinman@noao.edu}, C. Cacciari$^{2}$, 
A. Bragaglia$^{2}$, A. Buzzoni$^{2}$, A. Spagna$^{3}$ \\
\medskip \\
$^{1}$ NOAO, P.O.\ Box 26732, Tucson, AZ 85726-6732, 
USA\thanks{NOAO is operated by AURA Inc.\ under contract with the 
National Science Foundation.}\\
$^{2}$ INAF, Osservatorio Astronomico di Bologna, Via Ranzani 1, I-40127 
Bologna, Italy\\
$^{3}$ INAF, Osservatorio Astronomico di Torino,  Via Osservatorio 20, I-10025 
Pino Torinese, Italy}

\date{Received 2006 Novembre 28; Accepted 2006 December 11}

\pagerange{\pageref{firstpage}--\pageref{lastpage}} \pubyear{2006}

\maketitle

\label{firstpage}

\begin{abstract}

 Radial velocities and proper motions (derived from the GSC-II database) are
 given for 38 RR Lyrae (RRL) stars and 79 blue horizontal branch (BHB) stars 
 in a $\sim$ 200~deg$^2$ area around the North Galactic Pole (NGP). 
 Both heliocentric (UVW) and galactocentric (V$_{R}$, V$_{\phi}$, V$_{z}$) 
 space motions  are derived for these stars using a
 homogeneous distance scale consistent with $(m-M)_0$=18.52 for the LMC. 

 An analysis of the 26 RRL and 52 BHB stars whose height (Z) above the plane
 is less than 8 kpc, shows that this halo sample is {\em not homogeneous}.
 Our BHB sample (like that of Sirko et al. 2004b) has a zero galactic rotation
 (V$_{\phi}$)       and roughly isotropic velocity dispersions. The RRL sample
 shows a definite retrograde rotation (V$_{\phi}$ = $-95\pm$29 km~s$^{-1}$) 
 and non-isotropic velocity dispersions. The {\em combined} BHB and RRL sample
 has a retrograde galactic rotation (V) that is similar to that found by
 Majewski (1992) for his sample of subdwarfs in SA~57. The velocity
 dispersion of the RRL stars that have a positive W motion is significantly 
 smaller than the dispersion of those ``streaming down" with a negative W. 
 Also, the ratio of RRL to BHB stars is smaller for the sample that has  
 positive W. 

 Our halo sample occupies 10.4 cubic kpc at a mean height of 5 kpc above the
 Galactic plane. In this volume, one component (rich in RR Lyrae stars)
 shows retrograde rotation and the streaming motion that we associate with
 the accretion process. The other component (traced by the BHB stars) shows
 essentially no rotation and less evidence of streaming. These two 
 components have HB morphologies that suggest that they may be the field 
 star equivalents of the young and old halo globular clusters respectively.
 Clearly, it is very desirable to use more than one tracer in
 any kinematic analysis of the halo.

\end{abstract}


\section{Introduction}

 A well known view of the globular-cluster halo is that it consists of an Old
 Halo and a Young Halo whose clusters have different HB morphologies that
 can be interpreted as an age difference (Zinn 1993). The two systems are to
 some extent coterminous but the clusters whose 
 R$_{gal}$ $<$ 6 kpc all belong to the Old Halo, while the Young Halo 
 predominates in the outer parts of our Galaxy. 
 Recent analyses of cluster ages (Rosenberg et al. 1999; De Angeli et al.
 2005) show that the most metal-poor clusters are all older and coeval while the
 more metal-rich are younger and have a significant dispersion in age. The
 outer parts of the Galaxy show a somewhat larger dispersion in cluster age.
 The Old Halo system has a prograde galactic rotation while that of the
 Young Halo is retrograde.  The Old Halo clusters have a high ratio of
 blue horizontal branch (BHB) to RR Lyrae (RRL) stars. 
 Layden (1996) found that the field RRL stars with 
 3 $\leq$ R$_{gal}\leq$ 6 kpc were not kinematically as cool as the Old Halo
 clusters since they have a larger velocity dispersion and a milder prograde 
 galactic rotation. Layden also analysed the radial velocities of the RRL 
 stars with $|$Z$|$ $>$ 5 kpc and whose R$_{gal}$ projected on the plane were
 between 6 and 10 kpc; these showed a marginally retrograde rotation. He
 accounted for these results in terms of the preferential destruction of the
 redder-HB clusters in the inner halo and by these redder-HB clusters having
 a greater fraction of RRL  stars per unit luminosity.
 Borkova and Marsakov (2003) analyzed the space-motions of local RRL  stars
 and also found evidence for two metal-poor ([Fe/H]$<$$-$1.0) systems. One  
 (corresponding to the Old Halo) is 
 associated with blue-HB clusters and has a spherical distribution and a 
slightly prograde galactic rotation. The other (corresponding to the Young Halo)
 is more spatially extended, is
 associated with red-HB clusters and has stars on eccentric and retrograde 
  orbits. A correlation between the kinematics of globular clusters and their
  Oosterhoff type was pointed
 out by van den Bergh (1993) and also by Lee \& Carney (1999). They found that
 Oosterhoff type II clusters belong to an older system with prograde orbits
 while  Oosterhoff type I clusters to a younger (accreted) system on retrograde 
 orbits. Lee \& Carney associated the field RRL  stars with $|$Z$|$ $<$ 3
  kpc and $|$Z$|$ $>$ 5 kpc with Oosterhoff types II and I respectively. 
The discovery of the disintegrating Sgr dwarf galaxy (Ibata, Gilmore \& Irwin
 1994) gave very strong support to the idea that the outer field-star halo
 parallels the Young cluster halo and is 
 composed of the tidal debris from such accreted satellites. This has prompted
 searches for other satellites and consequent non-uniformities
 in the structure of the halo.

Surveys to discover such substructure include among others, a K-giant 
survey (Morrison et al. 2000), an APM carbon star survey 
(Totten \& Irwin 1998; Totten, Irwin \& Whitelock 2000; Ibata et al. 2001), 
and the Sloan Digital Sky Survey (SDSS) for A stars (Yanny et al. 2000) and 
RRL stars (Ivezi\'c et al. 2000). 
All have shown inhomogeneities in the distribution of stars in the outer
halo that are mostly debris from the Sgr dSph (Newberg et al. 2002). 
More recent work with SDSS data has shown that these "streams" and 
 overdensities are very extensive (Lupton et al. 2005, Belokurov et al. 2006).
Many of these surveys study the {\em distant} halo:  88\% of the 
large sample of BHB stars 
discovered  by Sirko et al. (2004a) in the SDSS lie beyond 10 kpc and only 4
of these stars lie within 6 kpc. Thus, since proper motions are 
largely lacking for these stars, what we know of their kinematics 
can only come from analyzing their radial velocities 
(Sirko et al. 2004b; Thom et al. 2005). 

The kinematics of the {\em local} halo is much better known because  
accurate proper motions are available  (Martin \& Morrison 1998; 
Maintz \& de Boer 2005).\footnote {Summaries of other work on the local 
halo are given in Kinman et al. (2003) and in Table 1 of Sirko et al.
(2004b).} The mean halo rotation in the solar neighbourhood is close to zero 
(i.e. $\langle V \rangle \sim$ --220 km s$^{-1}$) or is slightly prograde, 
and it is well established that the velocity dispersions in the local halo are 
non-isotropic
( $\sigma_{U} > \sigma_{V} > \sigma_{W}$). 
  Helmi et al. (1999) studied 97  
metal-deficient ([Fe/H]$\leq$$-$1.6) red-giants and RRL  stars within 1 kpc
of the Sun and found that about $\sim$10\% of these had space-motions that 
showed that they came from a single disrupted satellite. This has been confirmed
 with a larger sample (234 stars within 2.5 kpc of the Sun) by Kepley et al.
 (2006) who also found evidence for two other possible ``streams". 

One of the first indications of structure in the halo 
was the discovery of retrograde motion for a group of subdwarf halo
stars in Selected Area (SA) 57 at the North Galactic Pole (NGP)  by Majewski 
(1992). Later,  Majewski, Munn \& Hawley (1996) found that these halo stars were
clumped in phase-space and [Fe/H]; in particular, they found a retrograde group 
($\langle V\rangle = -275 \pm 16$~km\,s$^{-1}$) with [Fe/H] $<$ --0.8 moving 
towards the galactic plane. 
Kinman et al. (1996) also found a preponderance of negative radial velocities
(i.e. motion towards the plane) among BHB and RRL stars in the same area 
of the sky. A detailed analysis of the Sgr stream of debris using 2MASS 
M-giants (Majewski et al. 2003; Law, Johnson \& Majewski 2005) shows that 
the Sgr stream may 
be inflowing from the general direction of the NGP towards the solar 
neighbourhood. Mart\'{\i}nez-Delgardo et al. (2006) have identified the Sgr 
leading arm with the overdensity in Virgo and predict highly negative
radial velocities in the North Galactic Cap but not at the area of the Pole
itself. 
On the other hand, neither Carney (1999) nor Chiba \& Beers (2000) find 
retrograde motion among different samples of nearby halo stars whose
 Z$_{max}$ $\geq$ 4 kpc. Vallenari et al. (2006) used the Padova Galaxy model
 to analyse the proper motions in a 26.7 deg$^{2}$ near the NGP. Their halo 
 component (for Z $<$ 7 kpc) showed no significant rotation and had 
 non-isotropic 
  velocity dispersions. The colour distributions of their halo stars suggest
 that they are primarily redder ($(B-V)$ $\geq$ +0.5) giant stars. 
 Other reports of nearby halo substructure include  Chiba 
\& Mizutani (2004), Altmann, Catelan \& Zoccali (2005) and Meza et al. (2005)
 who  discuss the debris from the globular cluster $\omega$ Cen 
(probably the nucleus of a disrupted dwarf galaxy). The orbit discussed by 
Chiba \& Mizutani does not rise more than 4 kpc 
above the plane in the direction of the NGP and so this debris is probably not 
connected with the substructure observed by Majewski et al. (1996) but might
account for a halo overdensity of F and G stars observed 
by Gilmore, Wyse \& Norris (2002). 
Duffau et al. (2006) have found a clump of both RRL        
and BHB stars at about 20 kpc that is
 coincident with an overdensity of F-type stars that
were previously found by Newberg et al. (2002). 
Juri\'{c} et al. (2006) have  found a significant overdensity of  halo stars at 
7 to 8 kpc in Virgo although a search by Brown et al. (2004) for overdensities 
of BHB stars out to 9 kpc gave negative results.
 Clewley \& Kinman (2006) have identified two new clumps that contain both 
RRL  and BHB stars at distances of 8 and 9 kpc. There is therefore 
considerable evidence that structure exists in the nearby halo
but accurate proper motions and radial velocities are needed in order to 
establish its significance. 

This paper (a continuation of Kinman et al. 1996, 2003, 2005) derives space
motions for  117 BHB and RRL  stars in $\sim$200 deg$^{2}$
 in the direction of the NGP.  
The sample of BHB stars in SA 57 that comes from the Case Low-Dispersion 
Sky Survey (Sanduleak 1988) and that was discussed by Kinman, Suntzeff \& Kraft 
(1994) has been expanded to cover about six Palomar Schmidt Sky Survey fields 
(Fig. 1) using sources discussed in Sec. 2.1. The RRL stars in the
same area of the sky have also been included and in many cases re-observed.
A data-base containing full details of these stars is in preparation 
(Kinman et al. 2007). BHB and RRL stars continue to be recognized as prime
halo tracers; recent surveys for them include  Sirko et al. (2004a),
Vivas et al. (2001); Clewley et al. (2002, 2004, 2005); 
Brown et al. (2003, 2004, 2005); Christlieb, Beers \& Thom (2005). Red giants,
subgiants and subdwarfs are more broadly representive of the halo but can be 
more difficult to separate from their disk counterparts and estimates of their 
distances are commonly less exact than those of the HB stars.
A major reason for studying the halo in the galactic polar regions is that 
the confusion between disk and halo stars is minimized in these directions; 
this is particularly important for the RRL stars which have a significant
disk population. The galactic extinction is also minimal. Further, Kepley et
al. (2006) have shown that streams may be more easily identified by their
radial velocities at the poles. Much previous work has depended on analyses
of radial velocities alone. Surveys for distant halo stars tend to be made 
at high galactic latitudes well away from the apex and antapex of galactic
rotation; the galactic rotation component of the radial velocity is small
compared with the random velocity components in these directions. Consequently
halo galactic rotations are not well determined from the radial velocity alone.
At the galactic poles, the relation between the galactic velocity vectors U, V  
and W of a star 
and its distance, radial velocity and proper motion is 
particularly simple. The U and V motions depend essentially only on the
distance and proper motions while the W motion depends almost entirely on
the radial velocity. In principle therefore, an analysis of the errors
 is more straightforward in the polar regions (and also along the prime 
galactic meridian) than in other directions.
   
It follows that it is not only important to get the best possible distances,
radial velocities and proper motions, but have a realistic assessment of 
their uncertainties. We feel that the proper motions that we have used 
(derived from the GSC-II database) are the best currently available. We 
have estimated their errors from the proper motions of QSO in the same fields.
 Presumably, the best proper motions will 
come from astrometric space missions such as GAIA and SIM (see 
e.g. GAIA, Concept and Technology Study Report 2004, Table A.4), but it will be
 several more years before these are available.
The majority of our radial velocities are of high quality ($\pm$10 km/s); 
there are some of lower accuracy but these do not significantly affect
the conclusions of this study. 
Finally, the database itself is not complete and it is hoped that it can be 
continually updated. In particular, it would be valuable to
include the halo red giants in this region (Majewski 2005). Clearly 
this study is an ongoing project rather than one where definitive results 
can be obtained with the current data.

We present our sample of BHB and RRL  stars and the data used for this 
study in Sect. 2; we  discuss our analysis and results in 
Sect. 3, and  Sect. 4 contains the summary and our conclusions.

\section{The Data}

\subsection{Target Selection} 

\begin{figure}
 \includegraphics[width=6.2cm,angle=90]{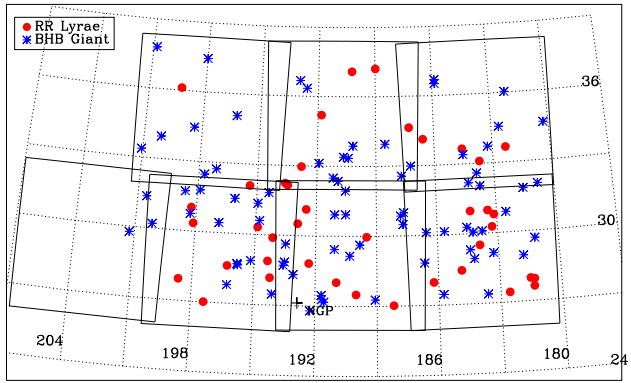}
\caption{The POSS-I fields 266, 267, 268 (top) and 320, 321, 322 and 323 
(bottom), used as first epoch plates to derive the proper motions of our 
target stars at the NGP. 
Right ascension, along the X axis, is expressed in arc degrees. 
The cross shows the position of the NGP. 
For comparison, the J2000 coordinates of the SA 57 are 13:08:38  +29:23:06   
(197.158  +29.385   in decimal degrees). 
}
\label{fig:fields}
\end{figure}

Our sample of halo tracers consists of 79 BHB and 38 RRL stars at distances
between 1.5 and 16 kpc located in an 
area of approximately 
$22^\circ\times12^\circ$ near the NGP.
This area, shown in Fig.~\ref{fig:fields}, is the combination of 7 POSS-I 
fields where proper motions from the GSC-II database were measured. 

The BHB stars of our sample were selected from the surveys of 
Sanduleak (1988),  Beers et al.~(1996), MacConnell et al.~(1993),  
and Kinman et al.~(1994). They are listed in Table \ref{t:bhb1}, 
with a simplified nomenclature: e.g we use 
 16549-51 for BPS BS 16549 0051 
(Beers et al. 1996), AF-003 for Case A-F 003 (Sanduleak 1988),  
and SA57-001 for SA 57 001 (Kinman et al. 1994). 
The single ID used in Table \ref{t:bhb1} is that of the earliest of these  
catalogues in which the star is mentioned;  alternative ID  will be given in 
Kinman et al.~(2007). 
These candidate BHB stars were confirmed by $uBV$-photometry (Kinman 
et al.~1994)  and spectroscopy. Most of the confirming spectra  
were taken  at the Kitt Peak 4-m Mayall telescope (Kinman et al.~1996).

The RRL stars (Table \ref{t:rrl1}) were mostly taken from the General 
Catalogue of Variable Stars (GCVS, Kholopov 1985) and subsequent Name-Lists, 
and have the traditional identification by constellation;
those without GCVS names  are taken from Kinman (2002). 
These are NSV5476, AF-791, AF-155, SA57-19, SA57-47, SA57-60 
and AF-882. The data for AF-031 and AF-042 are as yet unpublished. 

Intensity-weighted mean V magnitudes of the RRL stars are derived from 
recently observed light curves. 
Spectra of several RRL stars were obtained at the 3.5-m TNG telescope. 
Only 6 of these RRL stars and none of the BHB stars in our current sample 
are included in the catalogue of halo stars by Beers et al.~(2000).

K-magnitudes are available in the 2MASS catalogue for all our BHB and
RRL stars. The data from which the kinematic properties of these stars
were derived are given for the BHB and RRL stars in Tables \ref{t:bhb1} 
and \ref{t:rrl1} respectively. More detailed descriptions of this photometric 
and spectroscopic data will be given in Kinman et al. (2007).

\subsection{Reddenings, Absolute Magnitudes and Distances}

\subsubsection{Reddenings}

The reddening E$(B-V)$ was taken 
from  Schlegel et al. (1998).
The reddening corrections for other colours are derived from the 
 relations E$(V-K$ )=2.75E$(B-V)$, 
A$_V$=3.1E$(B-V)$ and A$_K$=0.35E$(B-V)$ (Cardelli et al. 1989).

\subsubsection{Absolute Magnitudes and Distances: RRL stars.}

\begin{figure}
\includegraphics[width=8.7cm]{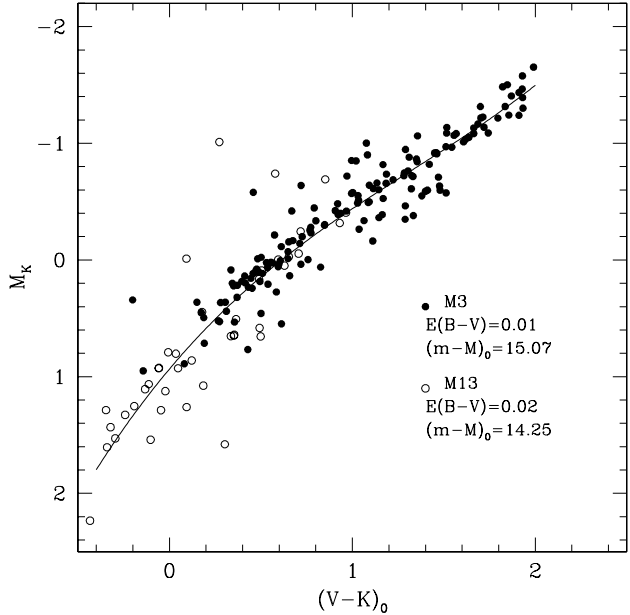} 
\caption{The combined HB of the globular clusters M3 and M13 in the IR plane 
M$_{K}$ vs (V--K)$_0$ (data from Valenti et al. 2004). The data have been 
corrected for reddening and shifted according to the individual distance moduli, 
as specified. The line shows the relation in Eq. (3). 
}
\label{f:mkhb}
\end{figure}

We consider three ways to estimate the absolute magnitudes and hence 
parallaxes of our RRL  stars:

\begin{enumerate}

\item 
 The visual absolute magnitude $M_{V}$ is given in terms of the metallicity
 [Fe/H] by the empirical relation:

\begin{equation}
 M_{V} = 0.214[Fe/H] + 0.86
\end{equation}
 
\noindent   
 We adopt the coefficients given by Clementini 
et al.~(2003) which imply that M$_{V}$ = 0.54 for [Fe/H] = $-$1.5.
They give an LMC modulus of $\sim$18.52 using the 
$\langle V_{0} \rangle$ = 19.064 that Clementini et al.~(2003) derive from 
their sample of about 100 RRL stars in the LMC bar.   
This M$_{V}$ estimate is therefore squarely in the middle of the range derived 
from  recent LMC distance determinations, most notably 18.50 (Freedman et 
al. 2001) and 18.54 (Tammann et al. 2003). The metallicity, [Fe/H], 
however, is known for only just over half of our RRL sample 
(see Table \ref{t:rrl1}). We assume [Fe/H] = --1.6 for the remaining 
stars since this is close to the mean metallicity of a halo population. 
An error of $\pm$0.5 dex in [Fe/H] leads to an error of $\pm$0.1 mag. in the 
distance modulus and about 5\% in the parallax. The parallaxes derived by
this method ($\Pi_{V}$) and their errors ($\sigma_{\Pi}$) are given in columns
8 and 9 respectively in Table \ref{t:rrl1}.

\item  
The infrared absolute magnitude $M_K$ is derived from the metallicity
and period. In the case of the type ab RRL stars we use the relation in the 
form given by Nemec et al. (1994):  

\begin{equation}
M_{K} =  -2.40 \log P + 0.06[Fe/H] -1.06
\end{equation}

\par\noindent
where the zero point is on the same scale as that given in (iii) below.
In the case of the c-type variables (indicated by an asterisk against their
$\log P$ in column 4 of Table \ref{t:rrl1}), the periods must be 
``fundamentalized" 
before using them in this relation. For this we have assumed that the ratio
of the first overtone (c-type) to fundamental (ab-type) period is 0.745 
(Clement et al. 2001). The parallaxes derived in this way ($\Pi_{K}$) are
given in column 11 of Table \ref{t:rrl1}.

\item 

 The infrared absolute magnitude $M_K$ is derived from the $(V-K)_0$ colour 
 using an empirical relation derived from data for non-variable stars in the 
globular clusters M3 and M13 (Valenti et al. 2004).
These two clusters have similar metallicities which are both close to the mean 
for the field halo population. We assumed reddenings E$(B-V)$ of 0.01 and 0.02 
and distance moduli $(m-M)_0$ of 15.07 and 14.25 for M3 and M13 respectively.  
These M3 and M13 data (Figure \ref{f:mkhb}) can be fitted by the 3-$\sigma$ 
rejection cubic polynomial:
\begin{eqnarray}
M_{K} = 0.878 - 1.812(V-K)_0 +0.675(V-K)_0^2  \nonumber\\ 
                                             -0.183(V-K)_0^3  
\end{eqnarray} 
This relation only holds for stars with the metallicity of M3 and M13. 
The  $M_{K}$ vs. \ $(V-K)_{0}$  relations for ZAHB stars of different [Fe/H]
were obtained from the models of VandenBergh et al. (2000). These models were
kindly transformed into  $M_{K}$ vs. \ $(V-K)_{0}$  by A. Sollima using the 
color-temperature relation given by Bessell et al. (1998). 
 The ZAHB relations  show that in the colour range of the RRL stars
an expression for $M_{K}$ needs a metallicity term $\sim$  $+$0.18[Fe/H]. 
  Equation (3) is therefore amended to:
\begin{eqnarray}
M_{K} = 1.166 +0.18[Fe/H] - 1.812(V-K)_0   \nonumber\\
          +0.675(V-K)_0^2   -0.183(V-K)_0^3  
\end{eqnarray} 

 The parallaxes ($\Pi_{HB}$),
 derived from equation (4), are given in column 12 
 of Table \ref{t:rrl1}. 

\end{enumerate} 

 The mean difference between the $M_K$ derived from equation (4) and that 
 from equation (2) for our 38 RRL stars is +0.025$\pm$0.022 mag with a 
 dispersion of 0.134 mag; this is satisfactorily small. Mean infrared 
 magnitudes $<K>$ were derived from the single epoch 2MASS magnitudes using
 the template $K$ light curves given by Jones et al. (1996) and
 the most appropriate ephemerides available. The $K$-amplitudes are such that 
 errors of $\sim$0.20 mag. in $<K>$ are possible if the estimated
 phases are in error by 0.1. 
The letter following $<K>$ in Table \ref{t:rrl1} gives the photometric 
quality \footnote{The letters A, B, C, D and U correspond to the observation 
having a S/N of $>$10, $>$7, $>$5, $<$5 and being an upper limit respectively.}
of the 2MASS data. Thirty five of the RRL sample have
 quality A ($\sigma_{K}$ $\leq$0.10 mag); for these $\Pi_{K}$/$\Pi_{V}$ = 
 1.010 $\pm$ 0.010 with a dispersion of 0.062. The twenty with no metallicity
 estimate (for which we assumed [Fe/H] = $-$1.6) have $\Pi_{K}$/$\Pi_{V}$ =
 1.022$\pm$ 0.014 and a dispersion of 0.064 while for 
 the fifteen variables for which [Fe/H] is known, 
 this ratio is 0.996$\pm$ 0.015 with a dispersion of 0.058. 
 The parallaxes derived from infrared magnitudes provide a useful 
 confirmation of the validity of the parallax ($\Pi_{V}$) derived from 
 visual magnitudes, but the greater uncertainty in the infrared magnitudes 
 compared with the visual magnitudes of our program stars and the greater
 uncertainties of 
 equations (2), (3) and (4) compared with  equation (1) has led us to 
 adopt ($\Pi_{V}$) alone for the calculation of the RRL distances.

\subsubsection{Absolute Magnitudes and Distances: BHB stars.}

 For the BHB stars, $M_V$ can be estimated from the $M_V$ vs.\ $(B-V)$ 
 relation given by Preston et al. (1991). We adjusted this relation
 so as to be consistent with the absolute scale that we have 
 adopted for the RRL stars, to get:  

\begin{eqnarray}
M_{V}=1.00 -4.423(B-V)_{0}+17.74(B-V)_{0}^{2}   \nonumber\\
                -35.73(B-V)_{0}^{3}  
\end{eqnarray} 

 The parallax derived from equation (5) is given as $\Pi_{HBP}$ in Table 
 \ref{t:bhb1}.
 Various problems involved with deriving $M_{V}$ from $(B-V)_0$  have been 
 discussed by Brown et al.~(2005). One concern is that the data for the 15 
 globular clusters on which this relation is based are now quite old.   
 For comparison, we therefore derived a similar cubic from more recent 
 data for the intermediate-metallicity clusters M3 (Ferraro et al. 1997) 
 and M13 (Paltrinieri et al. 1998):

\begin{eqnarray}
M_{V}=1.145 -5.615(B-V)_{0}+16.28(B-V)_{0}^{2}   \nonumber\\
                -14.93(B-V)_{0}^{3}  
\end{eqnarray} 

 The parallax derived from equation (6) is given as $\Pi_{HBC}$ in Table 
 \ref{t:bhb1}.
 The mean value of the ratio $\Pi_{HBC}$/$\Pi_{HBP}$ is 1.032$\pm$0.003 with
 a dispersion of 0.027. This ratio is adjusted to be unity at the blue edge of
 the instability strip \footnote{$(B-V)_0$$\sim$0.20~(Sandage 2006, Fig. 8).}
 and increases to 1.107 for the bluest BHB star 
 in our sample ($(B-V)_0$ = $-$0.08). We assume that the difference ($\Delta$)
 between these parallaxes is a measure of the likely systematic error in 
 $\Pi_{HBP}$. If $\sigma$ is the error in $\Pi_{HBP}$ (as deduced from the error
 in $(B-V)_0$) then its adopted error ($\sigma_{\Pi}$ given in column 9 of 
 Table \ref{t:bhb1}) is  $\surd(\Delta^2 + \sigma^2)$.

 Equation (3) can also be
 used to derive a  parallax for the BHB stars ($\Pi_{HVK}$ in
 column 10 of Table \ref{t:bhb1}).
 Fifty of them have 2MASS $K$ magnitudes of 
 quality A; for these the ratio $\Pi_{HVK}$/$\Pi_{HBP}$ = 0.950$\pm$0.005
 with a dispersion of 0.036. This ratio (adjusted to be unity at the blue
 edge of the instability strip)  decreases to 0.888 for the bluest star in
 this sample ($(B-V)_0$ = $-$0.02). Equation (3) is least-defined for its
 bluest colours since these correspond to the faintest HB stars in the 
 calibrating globular clusters. The derivation of parallaxes from 
 $(V-K)_0$ for these stars is likely to have significant advantages over the
 use of $(B-V)_0$, but currently we need better calibration and more accurate
 magnitudes for the fainter program stars before the method can be used 
 with confidence. Consequently only $\Pi_{HBP}$ is used in this paper.

\subsection{Radial velocities}

The radial velocities (RV) for our halo stars were either obtained from 
spectra obtained at the 4m-RC (KPNO) and 3.5m-LRS (TNG) spectrographs or from 
the literature. 
A more detailed discussion that includes the sources of these velocities and 
the individual galactocentric velocities will be given in Kinman et al. (2007).
The  heliocentric RV (with errors) of the program stars are listed in 
Tables \ref{t:bhb1} and \ref{t:rrl1}. We have no  RV for 7 of the RRL stars; 
these all have  $Z > 9$~kpc.

\begin{center}
\begin{table}
\caption{Mean galactocentric radial velocities $\langle$RV$_{gal}$$\rangle$ 
for our sample of NGP halo stars at different heights (Z) above the plane and 
Right Ascensions (RA).}
\label{t:rvgal}
\begin{tabular}{@{}ccccc@{}} 
\noalign{\smallskip}
\hline 
   Z   &   RA   &   N   &  $\langle$RV$_{gal}$$\rangle$  &  Dispersion    \\
 (kpc) & h:m     &      &  km\,s$^{-1}$ & km\,s$^{-1}$ \\
\noalign{\smallskip}
\hline 
\noalign{\smallskip}
 $<$ 4 & All     &   26  & $+$ 9.6$\pm$14.1  &  66.8$\pm$9.3    \\
 $>$ 4 & All     &   84  & $-$26.1$\pm$10.7  &  92.0$\pm$7.1    \\
 $>$ 4 &$<$13:00 &   61  & $-$17.4$\pm$12.3  &  91.7$\pm$8.3    \\
 $>$ 4 &$>$13:00 &   23  & $-$48.9$\pm$21.2  &  94.1$\pm$13.9    \\
\hline
\noalign{\smallskip}
\end{tabular}           
\end{table}
\end{center}

\begin{figure}
\includegraphics[width=8.2cm]{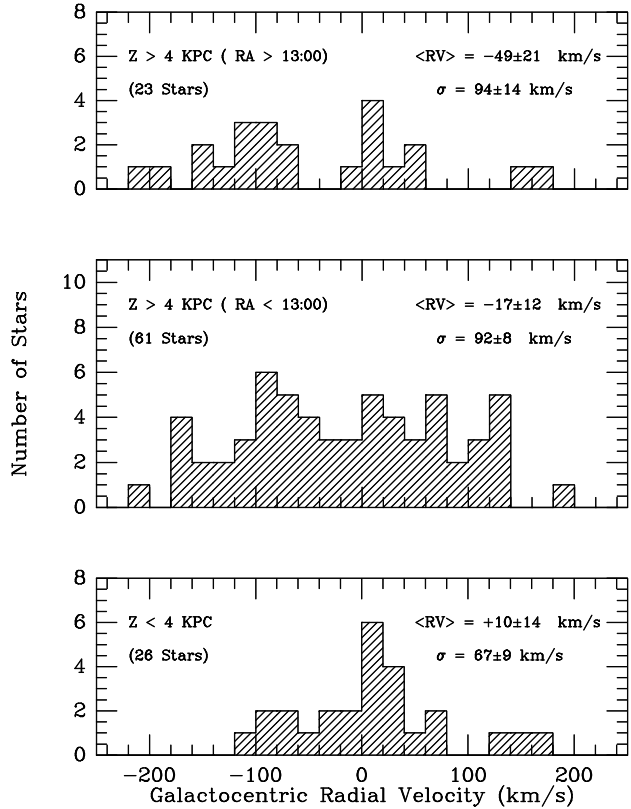}
\caption{The distributions of galactocentric radial velocity as a function of Z 
distance above the galactic plane, and at different Right Ascensions (RA).
}
\label{f:rvist3}
\end{figure}

%-------------------------------------------------------------------------

In this analysis,  we determine galactocentric radial velocities 
(RV$_{gal}$),
  by assuming a solar motion of
${\rm (U,V,W)_\odot=(10.0,\,5.25,\, 7.17)}$~km\,s$^{-1}$ with respect to the 
LSR  (Dehnen \& Binney 1998) and an LSR velocity of 220 km~s$^{-1}$. 
 The mean galactocentric radial velocities and their 
corrected dispersions\footnote{The method of correction is given in Sec. 3.1.}
are given in Table \ref{t:rvgal}; their distributions are shown in 
Fig. \ref{f:rvist3}.
We divide our sample of stars into those that lie below and above Z = 4 kpc; 
the group with Z $<$ 4 kpc 
does have a smaller dispersion, but the difference is  only significant at 
the 95\% level. 
There are eight stars in the Z $<$ 4 kpc sample that have a 
$\langle$RV$_{gal}$$\rangle$ of  +14.8 km\,s$^{-1}$ and a dispersion of only 
4.2  km\,s$^{-1}$;
a W-test (Shapiro \& Wilk 1965) of the whole Z$<$4 kpc sample, however, 
shows no departure from a Gaussian distribution. The dispersions found for
the sample with Z $>$ 4 kpc are in general agreement with those found for 
halo stars by Clewley et al. (2004) i.e. 108 $\pm$ 10 km\,s$^{-1}$ at distances 
11-52 kpc from the Sun, and by Sirko et al. (2004a) i.e. 
99.4$\pm$4.3 km\,s$^{-1}$ for their stars out to 30 kpc.
They are, however,  smaller than the dispersion of 120 km\,s$^{-1}$ (out to 
30 kpc) found by Battaglia et al. (2005). 

The velocity distribution should be Gaussian with a zero mean velocity 
(Harding et al. 2001) but our Z $>$ 4 kpc sample has a mean galactocentric RV
of $-$26.1$\pm$10.7 km\,s$^{-1}$ and an excess of large
negative velocities. This excess is most pronounced for the 23 stars 
whose  R.A. $>$ 13$^{h}$ and whose mean galactocentric RV is 
$-$49$\pm$21 km\,s$^{-1}$.
Individually the two samples with RA greater and less than 13$^{h}$:00$^{m}$
show no significant departure from a Gaussian distribution on a W-test 
(Shapiro \& Wilk 1965).

\subsection{Proper Motions and their Errors}

The proper motions  are derived from the plate material used for the 
construction of the GSC-II catalogue (Lasker et al. 1995; McLean et al. 2000).
We used multi-epoch positions that were derived from digitized Schmidt 
plates from the POSS-I, Quick-V, and POSS-II surveys (see Tables \ref{t:poss1} 
and \ref{t:poss2}). These cover a time-baseline of about 40 years and should
allow us to get a precision of a few mas per year.
We used  the procedure described by Spagna et al. (1996) to get the 
{\em relative} proper motions. 
 These were then transformed to the {\em absolute}
reference frame by forcing the extended extragalactic sources in the field to 
have no tangential motion.  The expected zero point accuracy is $\leq$ 1 mas yr$^{-1}$. 
The proper motions and their individual errors for our program objects 
are listed in Tables \ref{t:bhb1} and \ref{t:rrl1}; their formal 
r.m.s. errors are $\le 3$~mas\,yr$^{-1}$. Although these proper motions derived 
from the GSC-II are among the most accurate that are
available for the large-field surveys,  since an error of 
3 mas yr$^{-1}$ corresponds to 14.2 km s$^{-1}$ at 1 kpc, the errors in the
computed space motions of halo stars from their proper motions alone will be
comparable with the space motions themselves at a distance of $\sim$10~kpc.

\begin{center}
\begin{table}
\caption{Field centers of the POSS-I and Quick-V Schmidt plates
($6.4^{\circ} \times 6.4^\circ$) used in the proper motion
determinations.}
 \label{t:poss1}
%\begin{flushleft}  (103aO unfiltered and 103aE+red Plexiglass 2444)
\begin{tabular}{@{}cccccc@{}}
\noalign{\smallskip} \hline
Field   & RA & DEC & Epoch    & Epoch     & Epoch  \\
         & (deg)  & (deg)& POSS-I O$^a$ & POSS-I E$^b$ & Quick-V$^c$ \\
\hline \hline
%\multicolumn{4}{l}{POSS-I O  103aO unfiltered} \\
%       &       &      &      \\
 266 &183.82 &35.19 & 1956.350 & 1956.349 & 1983.133\\
 267 &190.76 &35.20 & 1950.365 & 1950.364 & 1983.133\\
 268 &197.69 &35.22 & 1950.367 & 1950.368 & 1983.133\\
 320 &183.83 &29.19 & 1955.293 & 1955.293 & 1983.293\\
 321 &190.28 &29.20 & 1950.275 & 1950.275 & 1983.293\\
 322 &196.73 &29.22 & 1955.288 & 1955.287 & 1982.386\\
 323 &203.18 &29.25 & 1950.428 & 1950.425 & 1982.387\\
\hline
\end{tabular}
$^a$ Photographic $O$ band (103aO unfiltered).\\
$^b$ Photographic $E$ band (103aE+red Plexiglass 2444).\\
$^c$ Photographic $V_{12}$ band (IIaD + Wratten 12).\\
\end{table}
\end{center}
%-----------------------------------------------------------

\begin{center}
\begin{table}
\caption{Field centers of the POSS-II Schmidt plates
($6.4^\circ\times 6.4^\circ$) that were used for the second epoch
positions in the proper motion determinations. }
 \label{t:poss2}
%\begin{flushleft}
\begin{tabular}{@{}cccccc@{}}
\noalign{\smallskip} \hline
Field   & RA & DEC & Epoch    & Epoch     & Epoch  \\
        & (deg)   & (deg)    & POSS-II J$^a$ & POSS-II F$^b$ & POSS-II 
N$^c$  \\
\hline \hline
 379  &   180.64 &  34.72  &   1996.216   & 1989.042 & 1998.373 \\
 380  &   186.62 &  34.72  &   1991.184   & 1991.184 & 1995.381 \\
 381  &   192.60 &  34.73  &   1990.211   & 1990.211 & 1995.384 \\
 382  &   198.58 &  34.74  &   1988.279   & 1989.026 & 1990.343 \\
 441  &   184.63 &  29.72  &   1993.285   & 1993.217 & 1996.312 \\
 442  &   190.36 &  29.73  &   1990.225   & 1990.225 & 1990.080 \\
 443  &   196.10 &  29.73  &   1995.233   & 1995.233 & 1993.288 \\
 444  &   201.83 &  29.74  &   1995.307   & 1995.307 & 1998.376 \\
\hline
\end{tabular}
 $^a$ Photographic $B_J$ band (IIIaJ+GG385).\\
$^b$  Photographic $R_F$ band (IIIaF+RG610).\\
$^c$  Photographic $I_N$ band (IV-N +RG9).\\
\end{table}
\end{center}

%----------------------------------------------------------- 

%\begin{landscape}
\begin{table*} 
\caption{Comparison of our proper motions (O) with those from the Hipparcos
(H) and Tycho (T) proper motions as given in the NOMAD catalogue 
Zacharias et al. (2004).                            
}
\label{t:err1}
\begin{tabular}{@{}rcccccccc@{}} 
\hline 
ID  & $\mu_{\alpha}$&$\mu_{\delta}$& Source 
& $\mu_{\alpha}$&$\mu_{\delta}$& Source 
& $\Delta$$\mu_{\alpha}$&$\Delta$ $\mu_{\delta}$ \\
(1) &(2)&(3) &(4) &(5) &(6) &(7) & (8) & (9)   \\
\hline 

  AF-113     & $-$25.9$\pm$1.9&$-$35.8$\pm$1.7&O&$-$26.3$\pm$0.9&$-$33.8$\pm$0.8&T&$+$0.4$\pm$2.1&$-$2.0$\pm$1.9  \\
 SA57-007    & $-$13.9$\pm$1.6&$-$28.5$\pm$2.2&O&$-$13.1$\pm$0.8&$-$24.1$\pm$1.1&T&$-$0.8$\pm$1.8&$-$4.4$\pm$2.5  \\
  AF-848     & $-$09.8$\pm$3.0&$-$18.0$\pm$2.8&O&$-$11.8$\pm$0.7&$-$20.5$\pm$0.9&T&$+$2.0$\pm$3.1&$+$2.5$\pm$2.9  \\
  AF-914     & $-$36.2$\pm$3.4&$-$15.0$\pm$3.4&O&$-$30.5$\pm$0.7&$-$13.1$\pm$0.7&T&$-$5.7$\pm$3.5&$-$1.9$\pm$3.5  \\
  S-Com      & $-$19.6$\pm$4.4&$-$18.4$\pm$3.9&O&$-$16.4$\pm$3.4&$-$16.5$\pm$2.2&H&$-$3.2$\pm$5.6&$-$1.9$\pm$4.5  \\
  U-Com      & $-$49.7$\pm$2.7&$-$16.4$\pm$2.1&O&$-$44.4$\pm$3.6&$-$13.8$\pm$2.0&H&$-$5.3$\pm$4.5&$-$2.6$\pm$2.9  \\
\hline 
\end{tabular}          
\end{table*}
%\end{landscape}

Proper motions for six of our brightest program stars are also given in the
Hipparcos and Tycho catalogues (available in the NOMAD catalogue, Zacharias
et al. 2004). These proper motions are compared in Table \ref{t:err1} 
where it is seen that the differences are largely within the expected errors.
These bright stars have large images on our survey plates and  
probably have the poorest proper motions in our survey. 
Consequently this comparison will give limited information on the systematic 
errors that may be present in our whole survey.
It is therefore crucial to have another independent estimate of the size of 
the proper motion errors. We therefore selected some 74 QSO brighter than 
18th magnitude that were listed in our fields by Hewitt \& Burbidge (1993) and
V\'{e}ron-Cetty \& V\'{e}ron (2001) and 62 type C (compact) objects from the
Kiso Schmidt Survey for UV-excess galaxies (Takase \& Miyauchi-Isobe 1993). 
These compact UV-excess galaxies have sufficiently ``stellar" images for 
them to be suitable for use as positional standards. As we noted previously
(Kinman et al. 2003),
some of these QSO  showed suprisingly large proper motions ($\geq$10 mas yr 
$^{-1}$) and spectra (kindly taken by Arjun Dey and Buell Jannuzi with the
Kitt Peak 4.0-m telescope) showed that five of them were stars. We therefore
rejected 15 of these objects whose proper motions (in either coordinate) were
more than 10 mas yr$^{-1}$. Table \ref{t:pmer} gives the mean proper motion 
in each field for the remaining 121 objects.
The fields are defined by the ID number of the POSS-II and POSS-I plates and 
some fields are labelled with an  ``a'' or ``b'' in order 
to identify partially overlapping POSS plates. Objects 
that fall on overlapping fields have multiple measures; these have been 
considered as separate measures. The unweighted mean of these 122 independent 
measures is given as ``ALL" in Table \ref{t:pmer} and is zero within its errors.
The r.m.s. uncertainty of this result is $\sim$0.3 mas yr$^{-1}$ in each 
coordinate.
Some 86\% of the QSO and compact galaxies and 81\% of the program stars are in
the five fields 381a-267, 382-268, 441-320, 442-321 and 443-322. The unweighted
means of the proper motions for these five fields are also zero within their 
errors. While we cannot be certain that there are not larger systematic errors 
in individual fields, the available data indicate that the r.m.s. systematic
error in the proper motion is $\sim$0.3 mas yr$^{-1}$ in each coordinate for
the data as a whole. The extragalactic standards are somewhat fainter on 
average than the program stars so that we must also be aware that 
magnitude-dependent errors may also be present.

\begin{table}
\caption{Proper motion systematic errors from QSO and Galaxies. 
The field identification in column 1 shows the ID numbers of the POSS-II and 
POSS-I plates.  
 Column 4 lists the 
number of QSO+Galaxies (Q+G) measures, and column 5 lists the number 
of BHB+RRL stars present in the corresponding fields.  } 
\label{t:pmer}
\begin{tabular}{@{}lcccc@{}} 
\noalign{\smallskip}
\hline 
Field  & $\mu_{\alpha}$ $\pm$ err. & $\mu_{\delta}$ $\pm$ err. &No. of&No. of\\
       & $ mas~yr^{-1}$   & $ mas~yr^{-1}$    & Q+G & stars \\
\noalign{\smallskip}
\hline 
\noalign{\smallskip}
379-266    & $\cdots$           & $\cdots$            &$\cdots$&  8\\
380a-267   &$-$3.0  $\pm \cdots$&$-$2.0  $\pm \cdots$ &  1&  5\\
380b-266   &$-$2.0  $\pm \cdots$&$+$4.0  $\pm \cdots$ &  1&  4\\
381a-267   &$-$1.68 $\pm$ 0.73  &$-$0.81 $\pm$ 1.35   &  9&  11\\
381b-268   &$+$0.27 $\pm$ 1.36  &$-$0.29 $\pm$ 1.15   &  7&  1\\
382-268    &$-$0.37 $\pm$ 0.87  &$+$0.49 $\pm$ 0.73   & 23&  9\\
441-320    &$-$0.24 $\pm$ 0.70  &$+$0.84 $\pm$ 0.95   & 14&  28\\
442-321    &$+$0.27 $\pm$ 0.50  &$-$0.66 $\pm$ 0.59   & 30&  26\\
443-322    &$+$0.16 $\pm$ 0.62  &$-$0.37 $\pm$ 0.69   & 29&  21\\
444a-322   & $\cdots$           &  $\cdots$           &$\cdots$&  1\\
444b-323   &$-$1.56 $\pm$ 0.74  &$+$0.78 $\pm$ 1.26   &  8&  3\\
\hline
ALL    &$-$0.25 $\pm$ 0.28  &$-$0.07 $\pm$ 0.31   &122&  117\\
FIVE FIELDS &$-$0.14 $\pm$ 0.31  &$-$0.14 $\pm$ 0.34   &105&  95\\
\noalign{\smallskip}
\hline 
\noalign{\smallskip}
\end{tabular}           
\end{table}
%----------------------------------------------------------- 

\section{Analysis and Results}

\subsection{The space-velocities U,V \& W}

The heliocentric space-velocity components U, V, and W were derived from 
 the  data listed in Tables \ref{t:bhb1} and \ref{t:rrl1}. We used the 
program by Johnson \& Soderblom (1987) (updated for the J2000 
reference frame and further updated with the transformation 
matrix derived from the Vol. 1 of the Hipparcos data catalogue).  
This program gives a right-handed system for U, V and W in which these vectors
are positive towards the directions of the Galactic centre, 
Galactic rotation and the NGP respectively\footnote{Other
authors (e.g. Majewski 1992) use a left-handed system in which U is positive
towards the Galactic anticentre.}. As noted in Sect. 2.2.2 and 2.2.3, 
$\Pi_{HBP}$ and $\Pi_{V}$ were adopted for the parallaxes of the BHB and RRL 
stars respectively.
No radial velocities are yet available for seven of the RRL stars. We therefore 
assumed, when calculating their  U,V \& W velocities, that their
radial velocity is zero with an error of 
 150~km\,s$^{-1}$. This should give acceptable U and V velocities since
these depend almost entirely on the proper motions; their W velocities must,
of course, be discarded.  These values of U,V \& W together with the height
(Z) of the star above the Galactic plane and the galactocentric distance 
(R$_{gal}$)  are given in Table \ref{t:bhb2} and 
Table \ref{t:rrl2} for the BHB and RRL stars respectively. 
We assume a solar galactocentric distance of 8.0 kpc.

\begin{center}

\begin{table*} 
\caption{The BHB stars of our sample. The U,V and W velocities (in km s$^{-1}$)
are heliocentric. The velocities V$_{r}$, V$_{\phi}$ and V$_{z}$ (in km s$^{-1}$) are in
galactocentric cylindrical coordinates. For further details, see text.
}
\label{t:bhb2}
\begin{tabular}{@{}rcccccccc@{}} 
\hline 
ID  &Z (kpc)  &R$_{gal}$ (kpc)&  U   &  V   &   W   & V$_{R}$  &  V$_{\phi}$ & V$_{z}$     \\
(1) &(2)&(3) &(4) &(5) &(6) &(7) & (8) & (9)   \\
\hline 

    16549-51 & 3.6 &  9.5 &$-$192$\pm$033&$-$246$\pm$026&$+$007$\pm$049&$-$183$\pm$033& $-$013$\pm$026&$-$014$\pm$049  \\
    AF-003   &14.8 & 18.5 &$+$199$\pm$296&$-$209$\pm$323&$-$039$\pm$071&$+$207$\pm$296& $+$028$\pm$323&$+$032$\pm$071   \\
    AF-006   & 3.2 &  9.1 &$+$011$\pm$048&$-$338$\pm$042&$+$013$\pm$013&$+$022$\pm$048& $-$106$\pm$042&$-$020$\pm$013   \\
    AF-011   &10.0 & 14.1 &$+$418$\pm$198&$-$534$\pm$277&$-$021$\pm$037&$+$432$\pm$198& $-$295$\pm$277&$+$014$\pm$037   \\
    AF-013   &15.6 & 18.9 &$-$130$\pm$153&$-$405$\pm$127&$+$084$\pm$048&$-$106$\pm$153& $-$183$\pm$127&$-$091$\pm$048   \\
    AF-727   & 4.3 &  9.9 &$+$048$\pm$033&$-$024$\pm$036&$-$012$\pm$012&$+$062$\pm$033& $+$206$\pm$036&$+$005$\pm$012   \\
    AF-022   &12.6 & 16.2 &$+$072$\pm$114&$-$370$\pm$082&$-$037$\pm$022&$+$085$\pm$114& $-$136$\pm$082&$+$030$\pm$022   \\
    AF-729   & 8.0 & 12.5 &$+$021$\pm$072&$-$331$\pm$074&$-$016$\pm$041&$+$027$\pm$072& $-$100$\pm$074&$+$009$\pm$041   \\
    AF-029   & 6.6 & 11.4 &$+$079$\pm$081&$-$105$\pm$064&$-$045$\pm$017&$+$090$\pm$081& $+$125$\pm$064&$+$038$\pm$017   \\
    AF-030   & 5.8 & 10.6 &$-$077$\pm$063&$-$324$\pm$059&$+$016$\pm$014&$-$065$\pm$063& $-$094$\pm$059&$-$023$\pm$014   \\
    16022-26 & 5.8 & 10.5 &$+$013$\pm$050&$-$288$\pm$065&$-$047$\pm$049&$+$025$\pm$050& $-$055$\pm$065&$+$040$\pm$049   \\
    AF-038   & 8.7 & 12.8 &$-$078$\pm$089&$-$423$\pm$127&$-$090$\pm$017&$-$070$\pm$089& $-$191$\pm$127&$+$083$\pm$017   \\
    AF-039   &12.2 & 15.7 &$-$280$\pm$091&$-$217$\pm$094&$+$011$\pm$041&$-$271$\pm$091& $+$006$\pm$094&$-$018$\pm$041   \\
    AF-041   & 7.4 & 11.8 &$-$148$\pm$058&$-$433$\pm$066&$+$077$\pm$013&$-$141$\pm$058& $-$200$\pm$066&$-$084$\pm$013   \\
    AF-045   & 3.4 &  9.1 &$-$055$\pm$019&$-$067$\pm$020&$+$007$\pm$010&$-$049$\pm$019& $+$164$\pm$020&$-$014$\pm$010   \\
    AF-046   &11.0 & 14.6 &$-$136$\pm$112&$-$192$\pm$162&$+$040$\pm$017&$-$128$\pm$112& $+$036$\pm$162&$-$047$\pm$017   \\
    AF-048   & 3.6 &  9.3 &$-$038$\pm$034&$-$211$\pm$037&$+$018$\pm$011&$-$029$\pm$034& $+$021$\pm$037&$-$025$\pm$011   \\
    AF-050   &15.2 & 18.3 &$+$042$\pm$216&$-$384$\pm$180&$+$019$\pm$050&$+$059$\pm$216& $-$149$\pm$180&$-$026$\pm$050   \\
    AF-052   & 6.0 & 10.7 &$-$116$\pm$110&$-$293$\pm$090&$-$088$\pm$018&$-$106$\pm$110& $-$062$\pm$090&$+$081$\pm$018   \\
    AF-053   & 6.8 & 11.4 &$+$148$\pm$059&$-$402$\pm$091&$+$047$\pm$015&$+$153$\pm$059& $-$173$\pm$091&$-$054$\pm$015   \\
    AF-063   &12.4 & 15.6 &$-$079$\pm$096&$-$316$\pm$075&$-$117$\pm$015&$-$069$\pm$096& $-$085$\pm$075&$+$110$\pm$015  \\
    16026-28 & 3.2 &  8.9 &$+$033$\pm$024&$-$055$\pm$022&$-$085$\pm$010&$+$038$\pm$024& $+$178$\pm$022&$+$078$\pm$010   \\
    AF-754   & 7.8 & 12.2 &$+$012$\pm$077&$-$252$\pm$120&$+$008$\pm$045&$+$019$\pm$077& $-$021$\pm$120&$-$015$\pm$045   \\
    AF-755   & 7.8 & 12.2 &$-$031$\pm$098&$-$296$\pm$113&$-$101$\pm$044&$-$027$\pm$098& $-$062$\pm$113&$+$094$\pm$044   \\
    AF-068   & 5.8 & 10.4 &$-$001$\pm$065&$-$195$\pm$052&$-$136$\pm$012&$+$008$\pm$065& $+$037$\pm$052&$+$129$\pm$012   \\
    AF-070   & 6.9 & 11.1 &$-$094$\pm$052&$-$231$\pm$066&$+$166$\pm$011&$-$085$\pm$052& $-$001$\pm$066&$-$173$\pm$011  \\
    AF-073   &16.8 & 19.6 &$+$233$\pm$125&$-$238$\pm$122&$-$001$\pm$042&$+$241$\pm$125& $-$023$\pm$122&$-$006$\pm$042  \\
    AF-075   & 8.3 & 12.1 &$+$231$\pm$103&$-$365$\pm$089&$-$020$\pm$014&$+$238$\pm$103& $-$137$\pm$089&$+$013$\pm$014  \\
    AF-076   & 9.9 & 13.3 &$+$065$\pm$102&$-$285$\pm$066&$+$046$\pm$013&$+$074$\pm$102& $-$054$\pm$066&$-$053$\pm$013  \\
    AF-077   & 9.2 & 12.9 &$+$175$\pm$146&$-$329$\pm$166&$-$126$\pm$022&$+$180$\pm$146& $-$104$\pm$166&$+$119$\pm$022  \\
    AF-078   & 4.4 &  9.5 &$+$118$\pm$027&$-$087$\pm$037&$-$165$\pm$010&$+$128$\pm$027& $+$144$\pm$037&$+$158$\pm$010  \\
    AF-769   & 3.5 &  9.1 &$-$116$\pm$036&$-$393$\pm$046&$+$021$\pm$011&$-$112$\pm$036& $-$158$\pm$046&$-$028$\pm$011  \\
    16026-67 & 2.9 &  8.6 &$-$041$\pm$037&$-$338$\pm$055&$-$040$\pm$010&$-$031$\pm$037& $-$106$\pm$055&$+$033$\pm$010  \\
    AF-100   & 3.9 &  9.1 &$-$181$\pm$032&$-$305$\pm$040&$+$049$\pm$010&$-$172$\pm$032& $-$072$\pm$040&$-$056$\pm$010  \\
    16466-08 & 6.3 & 10.7 &$-$123$\pm$076&$-$257$\pm$062&$+$025$\pm$040&$-$115$\pm$076& $-$018$\pm$062&$-$032$\pm$040  \\
    AF-108   & 3.8 &  9.0 &$-$072$\pm$032&$-$492$\pm$046&$-$069$\pm$010&$-$064$\pm$032& $-$260$\pm$046&$+$062$\pm$010  \\
    AF-111   & 5.0 &  9.8 &$-$282$\pm$047&$-$087$\pm$045&$-$037$\pm$011&$-$266$\pm$047& $+$157$\pm$045&$+$030$\pm$011  \\
    AF-112   & 2.6 &  8.6 &$+$069$\pm$023&$-$089$\pm$022&$+$009$\pm$010&$+$080$\pm$023& $+$142$\pm$022&$-$016$\pm$010  \\
    AF-113   & 1.6 &  8.3 &$-$025$\pm$014&$-$341$\pm$019&$+$020$\pm$010&$-$017$\pm$014& $-$109$\pm$019&$-$027$\pm$010  \\
    AF-115   & 8.1 & 11.9 &$-$017$\pm$060&$-$397$\pm$056&$+$041$\pm$011&$-$020$\pm$060& $-$164$\pm$056&$-$048$\pm$011  \\
   16466-15  & 2.9 &  8.7 &$-$052$\pm$042&$-$079$\pm$052&$+$146$\pm$011&$-$040$\pm$042& $+$154$\pm$052&$-$153$\pm$011  \\
    AF-130   &11.4 & 14.2 &$+$063$\pm$090&$-$209$\pm$068&$+$113$\pm$010&$+$073$\pm$090& $+$021$\pm$068&$-$120$\pm$010  \\
    AF-131   & 9.2 & 12.6 &$-$012$\pm$080&$-$314$\pm$078&$+$059$\pm$011&$-$007$\pm$080& $-$082$\pm$078&$-$066$\pm$011  \\
    AF-134   & 2.6 &  8.6 &$+$220$\pm$023&$-$204$\pm$019&$+$002$\pm$109&$+$230$\pm$023& $+$023$\pm$019&$-$009$\pm$010  \\
    16031-44 & 4.2 &  9.1 &$+$161$\pm$040&$-$203$\pm$056&$+$011$\pm$010&$+$170$\pm$040& $+$030$\pm$056&$-$018$\pm$010   \\
    15622-48 & 5.7 &  9.9 &$+$078$\pm$049&$-$279$\pm$064&$-$069$\pm$010&$+$087$\pm$049& $-$047$\pm$064&$+$062$\pm$010   \\
    15622-07 & 2.7 &  8.5 &$+$033$\pm$033&$-$172$\pm$039&$-$037$\pm$010&$+$042$\pm$033& $+$060$\pm$039&$+$030$\pm$010   \\
    AF-138   & 5.2 &  9.8 &$-$053$\pm$065&$-$325$\pm$059&$+$124$\pm$011&$-$049$\pm$065& $-$091$\pm$059&$-$131$\pm$011   \\
    15622-09 & 3.0 &  8.5 &$+$086$\pm$025&$-$178$\pm$032&$-$122$\pm$010&$+$095$\pm$025& $+$054$\pm$032&$+$115$\pm$010   \\
    AF-797   & 8.1 & 12.0 &$-$144$\pm$072&$-$350$\pm$050&$+$119$\pm$039&$-$149$\pm$072& $-$100$\pm$050&$-$126$\pm$039   \\
    AF-804   & 4.5 &  9.5 &$-$174$\pm$043&$-$243$\pm$034&$+$114$\pm$010&$-$165$\pm$043& $+$002$\pm$034&$-$121$\pm$010   \\
   SA57-001  & 5.7 &  9.9 &$+$228$\pm$069&$-$445$\pm$060&$-$113$\pm$010&$+$234$\pm$069& $-$216$\pm$060&$+$106$\pm$010   \\
   SA57-006  & 4.4 &  9.2 &$-$191$\pm$053&$-$257$\pm$050&$+$119$\pm$010&$-$183$\pm$053& $-$021$\pm$050&$-$126$\pm$010   \\
   SA57-007  & 1.6 &  8.2 &$+$030$\pm$013&$-$240$\pm$026&$-$084$\pm$049&$+$039$\pm$013& $-$008$\pm$026&$+$077$\pm$049   \\
   SA57-009  &11.8 & 14.3 &$-$169$\pm$136&$-$716$\pm$117&$+$067$\pm$010&$-$181$\pm$136& $-$477$\pm$117&$-$074$\pm$010   \\
   SA57-017  & 3.3 &  8.6 &$-$239$\pm$069&$-$356$\pm$063&$+$004$\pm$010&$-$231$\pm$069& $-$122$\pm$063&$-$011$\pm$010   \\
   SA57-021  &10.3 & 13.2 &$-$399$\pm$134&$+$066$\pm$146&$-$085$\pm$042&$-$358$\pm$134& $+$336$\pm$146&$+$078$\pm$042   \\
   SA57-029  &12.6 & 15.0 &$-$009$\pm$099&$-$307$\pm$105&$-$161$\pm$040&$-$008$\pm$099& $-$075$\pm$105&$+$154$\pm$040   \\
   SA57-032  & 7.8 & 11.3 &$-$074$\pm$069&$-$406$\pm$084&$+$017$\pm$012&$-$078$\pm$069& $-$169$\pm$084&$-$024$\pm$012   \\
   SA57-036  &11.9 & 14.3 &$+$003$\pm$123&$-$398$\pm$141&$+$012$\pm$011&$+$001$\pm$123& $-$166$\pm$141&$-$019$\pm$011   \\
\hline
\end{tabular}
\end{table*}

%\thispagestyle{empty}
%\topmargin 40mm

%\newpage 

\addtocounter{table}{-1} 
\begin{table*}
\caption{continued }
\label{t:bhb2}
\begin{tabular}{@{}rcccccccc@{}} 
\hline 
ID  &Z (kpc)  &R$_{gal}$ (kpc)&  U   &  V   &   W   & V$_{R}$  &  V$_{\phi}$ & V$_{z}$    \\
(1) &(2)&(3) &(4) &(5) &(6) &(7) & (8) & (9)   \\
\hline 
   SA57-040  & 7.4 & 10.8 &$-$033$\pm$049&$-$128$\pm$046&$+$036$\pm$010&$-$019$\pm$049& $+$105$\pm$046&$-$043$\pm$010   \\
   SA57-041  &13.0 & 15.1 &$-$124$\pm$115&$-$341$\pm$114&$+$146$\pm$011&$-$124$\pm$115& $-$099$\pm$114&$-$153$\pm$011   \\
   SA57-045  & 8.2 & 11.5 &$+$206$\pm$050&$-$242$\pm$045&$-$113$\pm$010&$+$213$\pm$050& $-$030$\pm$045&$+$106$\pm$010   \\
    AF-825   & 4.8 &  9.5 &$-$040$\pm$023&$-$191$\pm$030&$-$114$\pm$049&$-$027$\pm$023& $+$043$\pm$030&$+$107$\pm$049   \\
   SA57-046  &16.1 & 17.7 &$-$111$\pm$143&$-$483$\pm$157&$+$174$\pm$040&$-$123$\pm$143& $-$241$\pm$157&$-$181$\pm$040   \\
   SA57-055  &11.6 & 14.0 &$+$136$\pm$141&$-$315$\pm$102&$-$201$\pm$014&$+$134$\pm$141& $-$100$\pm$103&$+$194$\pm$014   \\
    AF-841   & 4.5 &  9.2 &$-$012$\pm$029&$-$273$\pm$035&$+$006$\pm$010&$-$006$\pm$029& $-$041$\pm$035&$-$013$\pm$010   \\
    AF-848   & 1.4 &  8.1 &$+$014$\pm$019&$-$144$\pm$021&$-$099$\pm$010&$+$025$\pm$019& $+$088$\pm$021&$+$092$\pm$010   \\
   SA57-066  & 6.1 & 10.0 &$-$009$\pm$087&$-$168$\pm$075&$-$225$\pm$013&$+$005$\pm$087& $+$064$\pm$075&$+$218$\pm$013   \\
    AF-854   & 6.9 & 10.9 &$-$295$\pm$058&$-$211$\pm$052&$+$019$\pm$049&$-$279$\pm$058& $+$067$\pm$052&$-$026$\pm$049   \\
   SA57-080  & 4.6 &  9.1 &$-$083$\pm$038&$-$192$\pm$048&$-$024$\pm$010&$-$072$\pm$038& $+$044$\pm$048&$+$017$\pm$010   \\
    AF-866   & 4.1 &  9.0 &$+$030$\pm$042&$-$245$\pm$050&$-$003$\pm$050&$+$038$\pm$042& $-$016$\pm$050&$-$004$\pm$050   \\
   SA57-087  &10.9 & 13.4 &$-$077$\pm$135&$-$308$\pm$135&$-$074$\pm$018&$-$079$\pm$135& $-$064$\pm$135&$+$067$\pm$018   \\
    AF-900   & 3.4 &  8.6 &$-$018$\pm$020&$-$215$\pm$025&$+$128$\pm$049&$-$008$\pm$020& $+$018$\pm$025&$-$135$\pm$049   \\
   SA57-111  &13.6 & 15.4 &$-$160$\pm$089&$-$448$\pm$094&$-$080$\pm$014&$-$191$\pm$089& $-$181$\pm$094&$+$073$\pm$014   \\
    AF-909   & 7.0 & 10.4 &$+$060$\pm$042&$-$214$\pm$043&$-$121$\pm$011&$+$071$\pm$042& $+$010$\pm$043&$+$114$\pm$011   \\
    AF-914   & 1.7 &  8.2 &$-$162$\pm$031&$-$249$\pm$037&$+$151$\pm$049&$-$154$\pm$031& $-$010$\pm$037&$-$158$\pm$049   \\
    AF-916   & 4.7 &  9.2 &$+$073$\pm$033&$-$130$\pm$036&$-$115$\pm$049&$+$091$\pm$033& $+$094$\pm$036&$+$108$\pm$049   \\
    AF-918   & 4.1 &  8.7 &$-$055$\pm$041&$-$188$\pm$039&$+$001$\pm$049&$-$043$\pm$041& $+$047$\pm$039&$-$008$\pm$049   \\
\hline 
\end{tabular}           
\end{table*}
\end{center}

%----------------------------------------------------------- 

%\centering
\begin{center}

\begin{table*} 
\caption{The RR stars of our sample. The U,V and W velocities (in km s$^{-1}$)
are heliocentric. The velocities V$_{R}$, V$_{\phi}$ and V$_{z}$ (in km s$^{-1}$) are in
 galactocentric cylindrical coordinates. For further details, see text.
}
\label{t:rrl2}
\begin{tabular}{@{}rcccccccc@{}} 
\hline 
ID  &Z (kpc)  &R$_{gal}$ (kpc)&  U   &  V   &   W   & V$_{R}$  &  V$_{\phi}$ & V$_{z}$     \\
(1) &(2)&(3) &(4) &(5) &(6) &(7) & (8) & (9)  \\
\hline 

  GR-Com     &13.9 & 17.3 &$-$035$\pm$095&$-$470$\pm$111&$\cdots   $   &$-$001$\pm$095& $-$239$\pm$111&$\cdots   $      \\
  GH-Com     & 8.9 & 13.0 &$+$113$\pm$076&$-$373$\pm$096&$\cdots   $   &$+$133$\pm$076& $-$131$\pm$096&$\cdots   $      \\
  IQ-Com     &10.4 & 14.2 &$-$144$\pm$086&$-$137$\pm$065&$\cdots   $   &$-$142$\pm$086& $+$084$\pm$065&$\cdots   $         \\
  NSV-5476   & 7.4 & 12.0 &$+$011$\pm$062&$-$332$\pm$051&$-$019$\pm$050&$+$019$\pm$062& $-$100$\pm$051&$+$012$\pm$050   \\
  V-Com      & 3.4 &  9.2 &$-$221$\pm$086&$-$133$\pm$153&$-$018$\pm$028&$-$215$\pm$086& $+$092$\pm$153&$+$011$\pm$028   \\
  CD-Com     & 9.4 & 13.4 &$-$233$\pm$068&$-$703$\pm$085&$-$257$\pm$014&$-$214$\pm$068& $-$475$\pm$085&$+$250$\pm$014   \\
  AF-031     & 9.5 & 13.4 &$-$133$\pm$072&$-$444$\pm$124&$-$202$\pm$050&$-$118$\pm$072& $-$215$\pm$124&$+$195$\pm$050   \\
  TU-Com     & 4.4 &  9.7 &$+$194$\pm$095&$-$368$\pm$168&$-$074$\pm$016&$+$204$\pm$095& $-$134$\pm$168&$+$067$\pm$016   \\
  CK-Com     & 6.6 & 11.3 &$+$047$\pm$042&$-$557$\pm$049&$-$070$\pm$012&$+$051$\pm$042& $-$326$\pm$049&$+$063$\pm$012   \\
  AF-042     &14.1 & 17.3 &$-$099$\pm$151&$-$363$\pm$141&$\cdots   $   &$-$083$\pm$151& $-$135$\pm$141&$\cdots   $      \\
  CL-Com     & 7.7 & 11.9 &$+$076$\pm$083&$-$207$\pm$050&$-$025$\pm$016&$+$085$\pm$083& $+$026$\pm$050&$+$018$\pm$016   \\
  AT-CVn     & 6.8 & 11.4 &$-$089$\pm$086&$-$266$\pm$121&$+$089$\pm$052&$-$081$\pm$086& $-$032$\pm$121&$-$096$\pm$052   \\
  GY-Com     &11.4 & 14.8 &$+$293$\pm$089&$-$204$\pm$095&$\cdots   $   &$+$300$\pm$089& $+$044$\pm$095&$\cdots   $      \\
  GS-Com     &14.8 & 17.6 &$+$111$\pm$186&$-$335$\pm$293&$\cdots   $   &$+$126$\pm$187& $-$095$\pm$293&$\cdots   $      \\
  RR-CVn     & 2.6 &  8.8 &$-$053$\pm$031&$-$390$\pm$040&$+$014$\pm$011&$-$047$\pm$031& $-$157$\pm$040&$-$021$\pm$011   \\
  S-Com      & 1.7 &  8.3 &$-$055$\pm$034&$-$205$\pm$033&$-$066$\pm$004&$-$046$\pm$034& $+$027$\pm$033&$+$059$\pm$004   \\
  SV-CVn     & 2.7 &  8.8 &$+$090$\pm$018&$-$326$\pm$028&$+$081$\pm$028&$+$095$\pm$018& $-$098$\pm$028&$-$088$\pm$028   \\
  FV-Com     & 6.5 & 10.7 &$+$180$\pm$060&$-$154$\pm$061&$-$171$\pm$050&$+$190$\pm$060& $+$076$\pm$061&$+$164$\pm$050   \\
  U-Com      & 1.7 &  8.2 &$-$264$\pm$020&$-$314$\pm$019&$-$038$\pm$003&$-$255$\pm$020& $-$083$\pm$019&$+$031$\pm$003   \\
  SW-CVn     & 2.8 &  8.8 &$-$157$\pm$036&$-$200$\pm$036&$-$012$\pm$021&$-$146$\pm$036& $+$038$\pm$036&$+$005$\pm$021   \\
  DV-Com     & 7.3 & 11.0 &$+$280$\pm$066&$-$316$\pm$086&$-$127$\pm$010&$+$289$\pm$066& $-$083$\pm$086&$+$120$\pm$010   \\
  AF-791     & 6.9 & 11.0 &$-$362$\pm$061&$-$036$\pm$054&$-$230$\pm$012&$-$334$\pm$061& $+$226$\pm$054&$+$223$\pm$012   \\
  AW-Com     &10.5 & 13.4 &$+$235$\pm$072&$-$145$\pm$092&$\cdots   $   &$+$246$\pm$072& $+$080$\pm$092&$\cdots   $      \\
  TX-Com     & 5.6 & 10.0 &$+$236$\pm$052&$-$357$\pm$048&$-$088$\pm$049&$+$240$\pm$052& $-$134$\pm$048&$+$081$\pm$049   \\
  AP-CVn     & 4.6 &  9.5 &$-$181$\pm$038&$-$247$\pm$044&$-$009$\pm$049&$-$173$\pm$038& $-$007$\pm$044&$+$002$\pm$049   \\
  EM-Com     & 9.5 & 12.6 &$+$152$\pm$102&$-$334$\pm$087&$-$105$\pm$010&$+$155$\pm$102& $-$111$\pm$087&$+$098$\pm$010   \\
  AF-155     & 7.8 & 11.5 &$+$018$\pm$083&$-$262$\pm$090&$-$098$\pm$032&$+$025$\pm$083& $-$032$\pm$090&$+$091$\pm$032   \\
  TY-CVn     & 4.0 &  9.1 &$-$359$\pm$048&$-$303$\pm$058&$+$084$\pm$016&$-$353$\pm$048& $-$057$\pm$058&$-$091$\pm$016   \\
  IP-Com     & 7.2 & 10.8 &$+$157$\pm$087&$-$471$\pm$100&$-$111$\pm$030&$+$155$\pm$087& $-$246$\pm$100&$+$104$\pm$030   \\
 SA57-19     & 6.4 & 10.2 &$-$110$\pm$049&$-$364$\pm$052&$+$059$\pm$050&$-$104$\pm$049& $-$130$\pm$052&$-$066$\pm$050   \\
  EO-Com     & 6.9 & 10.6 &$-$154$\pm$078&$-$459$\pm$064&$+$089$\pm$010&$-$153$\pm$078& $-$222$\pm$064&$-$096$\pm$010   \\
  UV-Com     &10.1 & 12.9 &$-$293$\pm$067&$-$314$\pm$084&$+$076$\pm$011&$-$290$\pm$067& $-$059$\pm$084&$-$083$\pm$011   \\
  TZ-CVn     & 5.3 &  9.7 &$+$165$\pm$075&$-$361$\pm$074&$-$131$\pm$050&$+$166$\pm$075& $-$139$\pm$074&$+$124$\pm$050   \\
  SA57-47    & 5.9 &  9.8 &$+$137$\pm$054&$-$701$\pm$067&$-$134$\pm$037&$+$128$\pm$054& $-$474$\pm$067&$+$127$\pm$037   \\
  SA57-60    & 5.4 &  9.4 &$-$020$\pm$058&$-$473$\pm$068&$-$090$\pm$050&$-$017$\pm$058& $-$241$\pm$068&$+$083$\pm$050   \\
  EW-Com     & 8.0 & 11.2 &$-$220$\pm$063&$-$239$\pm$074&$+$050$\pm$012&$-$211$\pm$063& $+$015$\pm$074&$-$057$\pm$012   \\
  IS-Com     & 4.4 &  8.9 &$-$047$\pm$032&$-$440$\pm$044&$-$132$\pm$010&$-$045$\pm$032& $-$207$\pm$044&$+$125$\pm$010   \\
  AF-882     & 6.3 & 10.3 &$+$043$\pm$047&$-$286$\pm$050&$-$071$\pm$050&$+$044$\pm$047& $-$061$\pm$050&$+$064$\pm$050   \\
\hline 
\end{tabular}          
\end{table*}
\end{center}

\begin{table*}
\caption{Mean heliocentric space-velocities U, V \& W (km s$^{-1}$) for our program 
 BHB \& RRL stars as a function of height Z (kpc) above the galactic plane
and for local samples of these stars. Local~A is a sample of BHB stars 
 (Kinman et al. 2000) and Local~B and C are the HALO1 and HALO2 samples of RRL
 stars (Martin \& Morrison, 1998). Local D is the HALO2 sample after
 normalization to our distance scale.}
\label{t:uvwa}
\begin{center}
\leavevmode
\begin{tabular}[h]{cccccccccc} 
\noalign{\smallskip}
\hline 
\noalign{\smallskip}
Range & $<$Z$>$ & n$_{uv}$ &
  $<$U$>$  &   $\sigma_{u}$ &
  $<$V$>$  &   $\sigma_{v}$ &
   n$_{w}$      &
  $<$W$>$  &   $\sigma_{w}$  \\
in Z   &     &    &    &  &  &  &   &  &   \\  
 (1) & (2) & (3)& (4)& (5) &(6) & (7) & (8)& (9) &(10) \\  
\noalign{\smallskip}
\hline 
\noalign{\smallskip}
 0 $-$ 4 & 2.8 & 26 & $-$52$\pm$23&            110$\pm$15 &$-$242$\pm$22&       
103$\pm$14 & 26& 0 $\pm$14 &             67$\pm$ 9  \\
     &      &    & $(-$55$\pm$20)&              &($-$239$\pm$20)&               
  &  &($-$1$\pm$20) &              \\
    &  &  &  &  &  &  &  &  &     \\
 4 $-$ 8 & 6.0 & 52 & $-$12$\pm$20&            131$\pm$13 &$-$282$\pm$18&       
 111$\pm$11 & 52& $-$37$\pm$13 &            85$\pm$ 8  \\
     &      &    & $(-$7$\pm$16)&               &($-$278$\pm$14)&               
  &  &($-$37$\pm$11) &              \\
    &  &  &  &  &  &  &  &  &            \\
 8 $-$12 & 9.8 & 25 & $+$1$\pm$41&            172$\pm$24 &$-$328$\pm$34&        
 119$\pm$17 & 21& $-$32$\pm$24 &           106$\pm$15  \\
     &      &    & $(+$7$\pm$36)&               &($-$328$\pm$27)&               
  &  &($-$28$\pm$22) &               \\
    &  &  &  &  &  &  &  &  &     \\
 $\geq$12&14.1 & 14 & $-$26$\pm$40&            $\cdots$     &$-$349$\pm$24&     
 $\cdots$    & 11&     0$\pm$33 &            96$\pm$22  \\
     &      &    & $(-$24$\pm$36)&               &($-$350$\pm$23)&              
  &  &($-$2 $\pm$24) &               \\
    &  &  &  &  &  &  &  &  &     \\
\hline 
Local A  &  0    & 27 & $-$1$\pm$25& 129$\pm$18 &$-$210$\pm$15
       &79$\pm$11  & 27&$-$14$\pm$20 &101$\pm$14 \\
          &  &  &  &  &  &  &  &  &   \\
Local B  &  0    & 81 & $-$8$\pm$20&180$\pm$14 &$-$197$\pm$12
       &111$\pm$9   & 81&$-$8$\pm$10 &  93$\pm$7  \\
      &  &  &  &  &  &  &  &  &   \\
Local C  &  0    & 84 & $+$1$\pm$21&193$\pm$15 &$-$219$\pm$10
       &91$\pm$7   & 84&$-$5$\pm$10 &  96$\pm$7  \\
      &  &  &  &  &  &  &  &  &   \\
Local D  &  0    & 84 & $+$3$\pm$23&214$\pm$17 &$-$240$\pm$11
       &100$\pm$8   & 84&$-$8$\pm$11 & 108$\pm$8  \\
      &  &  &  &  &  &  &  &  &   \\

\noalign{\smallskip}
\hline 
\end{tabular}\end{center}\end{table*}

Table \ref{t:uvwa} gives the mean values $<$U$>$, $<$V$>$ and $<$W$>$ of 
these space-velocities for various ranges of Z. The mean values and  
dispersion given in parentheses were obtained by trimming the 10\% of the
sample that have the most extreme values. This has little effect on the mean
values and will not be discussed further. 
 The {\em rms} dispersions in these quantities 
 must be corrected for the errors in the individual U,V \& W 
space-velocities (given in Tables \ref{t:bhb2} and \ref{t:rrl2}). 
Following Jones \& Walker (1988), if the observed dispersion in U is 
Disp(U), and 
$\xi_{i}$ is the error in U of star $i$, then the corrected dispersion 
$\sigma_{u}$ is given by: 

\begin{equation}
\sigma_u^2 = (Disp(U))^2 - \frac{1}{n} \sum_{i=1}^{n} \xi_{i}^{2}  
\end{equation}

\noindent and similarly for the space-velocities V and W. These corrected
dispersions are given in columns (5), (7) and (10) of Table \ref {t:uvwa}.
For a given proper motion error, the space-velocity error $\xi_{i}$ increases
with distance; when it becomes comparable with the intrinsic velocity 
dispersion (at Z$\sim$10 kpc), the correction becomes unreliable. 
 Table \ref{t:uvwa} also gives $<$U$>$,$<$V$>$ \& $<$W$>$ and their 
dispersions for three halo samples from the solar neighbourhood. Local A is 
a sample of stars that were confirmed as BHB stars by high-resolution 
spectroscopy  (Kinman et al. 2000); these stars have Hipparcos proper motions 
and radial velocities with errors of a few km s$^{-1}$; their parallaxes and 
 space-velocities were determined in exactly the same way as for 
our program BHB stars. Local B is the HALO1 sample of local RRL  stars
(with [Fe/H] $<$ $-$1.3) from Martin \& Morrison (1998). Local C is 
their HALO2 sample in which the total space-velocity (which is defined by
equation (8) in Sec. 3.4) 
was used to exclude thick disk stars (as shown in our Fig. \ref{f:TSVZ}).
 Martin \& Morrison used M$_{V}$ = +0.73 
(at [Fe/H] = $-$ 1.9) to determine the distances of their RRL stars; they 
 plot the U, V \& W and their dispersions for their  HALO2
sample as a function of M$_{V}$ in Figs. 6 \& 7 respectively of {\em their 
 paper}. We used these plots to derive 
the space-motions of their HALO2 sample for our assumed M$_{V}$ of +0.54 at 
[Fe/H] = $-$1.5; these are given as Local D.  Changes in M$_{V}$ have the 
greatest effect on the V space-velocity and the U velocity dispersion in these
{\em local} halo samples.

If we assume that $<$V$_{LSR}>$  = $-$220 km s$^{-1}$ for zero halo rotation,
the corresponding $<$ V$_{hel}>$  will be $\sim$$-$225 km s$^{-1}$ 
(see Sect. 2.3).
Older {\em local} halo samples (see Table 5 of Martin \& Morrison) 
show  slightly prograde $<$V$>$. It is not clear how much these 
prograde  $<$V$>$ are a result of the adopted distance scales and/or the 
presence of disk stars in these halo samples.
 Most {\em local} samples have the problem that
either (a) they may contain some thick-disk stars (as in the case of the
 HALO1) or (b) may have some kinematic bias which is introduced in the attempt 
to remove the disk component (e.g. HALO2) or which is present in their 
original selection (e.g. Local~A). The differences between the space-motions
of the different local samples suggests that these putative systematic sampling 
errors are comparable in size to their statistical errors. 

\subsection{The Halo Rotation $<$V$>$}

\begin{figure*}
\includegraphics[width=10.0cm]{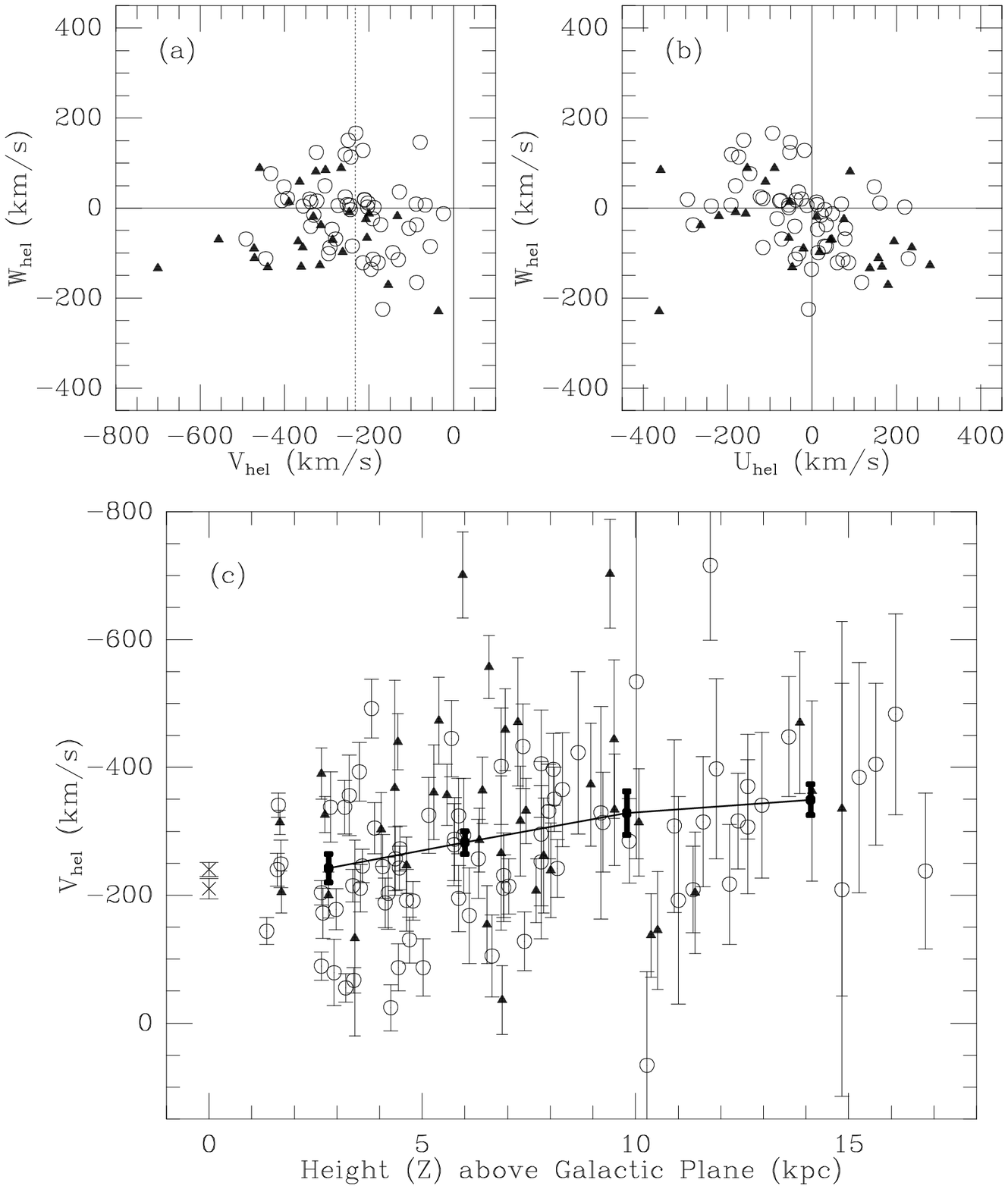} 
\caption{Plots of  V$_{hel}$ {\em vs} W$_{hel}$  and U$_{hel}$ {\em vs} 
W$_{hel}$ are shown in (a) and (b) respectively for the stars with 
Z $<$ 8 kpc. The vertical dotted line in (a) divides stars with retrograde 
orbits (on the left) from those with prograde orbits. The lower figure (c)
plots V$_{hel}$ {\em vs} the height (Z) above the galactic plane for the 
whole sample. The locations of the mean V$_{hel}$ of Local A and Local D are
shown by crosses. Filled triangles and open circles represent the
 RRL and BHB program stars respectively; the line connects the values of
$<$V$>$ for these stars. 
}
\label{f:UVW}
\end{figure*}

We see in Table \ref {t:uvwa} that the mean value $<$V$>$ of our halo sample
 becomes increasingly
{\em negative} (corresponding to more retrograde orbits) with increasing
height Z above the galactic plane. This change in $<$V$>$ is several times its
calculated error and so is formally significant. The V of the individual stars
are shown as a function of Z in Fig. \ref{f:UVW}(c); a line connects the
$<$V$>$ in this plot. The halo stars with 0 $<$ Z $<$ 4 kpc ($<$Z$>$ = 2.8 kpc)
have $<$V$>$ = $-$242$\pm$22 km s$^{-1}$. 
This is compatible with them having (on average) zero rotation;
 it is those with 4$<$Z$<$8 kpc  that have an excess of retrograde orbits.
Fig. \ref{f:UVW}(a) and (b) include only stars with Z $<$ 8 kpc and are to be
compared with stars in the same Z-range shown in Fig. 1 (c) and (b) 
respectively in Majewski, Munn \& Hawley (1996)(MMH) which are in the SA~57 at
the NGP.  The vertical dotted line 
in Fig. \ref{f:UVW}(a) and MMH (c) is the demarcation between prograde and
retrograde orbits. Both plots show a number of very  retrograde orbits with 
negative W velocities. The pronounced correlation between U and W in MMH (b) 
is present to a lesser extent in Fig. \ref{f:UVW}(b) \footnote{The 
different sign convention for U used by MMH causes negative U to correlate with
negative W in their case and positive W in our case.}. There is therefore 
qualitative agreement with MMH that there is structure in the phase-space 
distribution of these NGP stars. We test this quantitatively in the next 
Section.

\subsection{Is our NGP halo sample homogeneous?}  

 Table \ref {t:uvwb}  gives $<$U$>$ and $<$V$>$ and their corrected 
 dispersions $\sigma_{u}$ and $\sigma_{v}$ for our 78 halo stars with 
Z$<$ 8 kpc. Separate solutions are given for the RRL and BHB stars and also
for the stars with positive and negative W space-velocities. These show that:

\begin{table*}
\caption{Mean heliocentric space-velocities $<$U$>$ \&  $<$V$>$  
and the total space-velocity $<$T$>$  (km s$^{-1}$) for our program 
 BHB \& RRL stars for different ranges of the space-velocity W.}
\label{t:uvwb}
\begin{center}
\leavevmode
\begin{tabular}[h]{ccccccccc} 
\noalign{\smallskip}
\hline 
\noalign{\smallskip}
Star  &Range &  n$_{uv}$ &
  $<$U$>$  &   $\sigma_{u}$ &
  $<$V$>$  &   $\sigma_{v}$ &
  $<$T$>$  &   $\sigma_{T}$ \\ 
Type   & in W  &    &    &  &  &  &  &   \\  
 (1) & (2) & (3)& (4)& (5) &(6) & (7) &  (8) & (9)  \\  
\noalign{\smallskip}
\hline 
\noalign{\smallskip}
 RRL & W$<$ 0 & 20 & $+$12$\pm$41& 166$\pm$26 &$-$321$\pm$36& 136$\pm$ 21 &
   403$\pm$29  & 125$\pm$ 19  \\
 BHB & W$<$ 0 & 24 & $+$11$\pm$20&  78$\pm$11 &$-$216$\pm$24&  85$\pm$ 12 &
   272$\pm$22  & 104$\pm$ 15  \\
 All & W$<$ 0 & 44 & $+$12$\pm$21& 124$\pm$13 &$-$264$\pm$22& 128$\pm$ 14 & 
   332$\pm$20  & 131$\pm$ 14  \\
      &  &  &  &  &  & &   &    \\
 RRL & W$>$ 0 &  6 & $-$112$\pm$41& 135$\pm$39 &$-$351$\pm$31& 13$\pm$ 4  &
   410$\pm$31  &  69$\pm$ 20  \\
 BHB & W$>$ 0 & 28 & $-$64$\pm$20& 108$\pm$14 &$-$258$\pm$19&  82$\pm$ 11 &
   309$\pm$19  &  96$\pm$ 13  \\
 All & W$>$ 0 & 34 & $-$72$\pm$21& 112$\pm$14 &$-$275$\pm$17& 81$\pm$ 10  & 
   327$\pm$17  &  99$\pm$ 12  \\
      &  &  &  &  &  &  &   &   \\
 RRL &All W  & 26 & $-$17$\pm$35& 165$\pm$23 &$-$328$\pm$28& 118$\pm$ 16  &
   405$\pm$23  & 113$\pm$ 16  \\
 BHB &All W  & 52 & $-$29$\pm$16& 102$\pm$10 &$-$239$\pm$15&  93$\pm$ 9   &
   292$\pm$14  & 101$\pm$ 10  \\
 All &All W  & 78 & $-$25$\pm$16& 125$\pm$10 &$-$268$\pm$14& 109$\pm$ 9   &
   329$\pm$13  & 117$\pm$ 9  \\
      &  &  &  &  &  &  &  &    \\
 RRL(P$>$0.$^{d}$6) &All W  &  8 & $-$8$\pm$66& 163$\pm$41 &$-$414$\pm$75& 189$\pm$ 47  &
   490$\pm$50  & 132$\pm$ 33  \\
 RRL(P$<$0.$^{d}$6) &All W  & 18 & $-$21$\pm$44& 171$\pm$28 &$-$290$\pm$21& 33$\pm$ 6  &
   374$\pm$20  & 82$\pm$ 14  \\
      &  &  &  &  &  &  &  &    \\

\noalign{\smallskip}
\hline 
\end{tabular}\end{center}\end{table*}

\begin{figure}
\includegraphics[width=8.2cm]{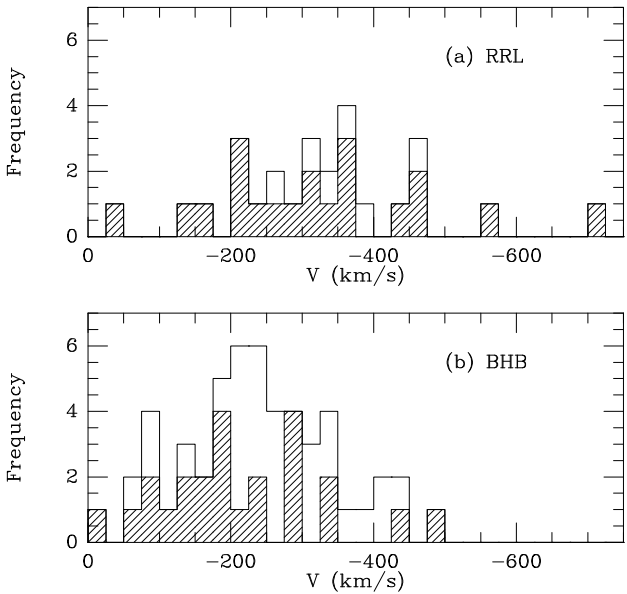}  
\caption{The distributions of V for (a) the RRL stars and (b) the BHB stars 
 in our NGP sample. Stars with hatched histograms are those with negative
 W velocities.
}
\label{fig:Vhist}
\end{figure}

\begin{itemize}

\item 

(a)  The combined solution for $<$V$>$ for both RRL and BHB stars and all W
 ($-$268$\pm$14 km s$^{-1}$) is in good agreement with that found by Majewski
 (1992) ($-$275 km s$^{-1}$). The $<$V$>$ of the BHB stars alone ($-$239$\pm$15
 km s$^{-1}$) is only mildly retrograde and compatible with zero halo 
 rotation. The RRL sample alone, however, is strongly retrograde ($-$328$\pm$28
 km s$^{-1}$). The dispersion $\sigma_{v}$ of the 20  RRL stars with negative W 
 (136$\pm$21 km s$^{-1}$) is much greater than that 
 (13$\pm$4 km s$^{-1}$) of the 6 RRL that have positive W. Despite the rather
 small numbers, a variance ratio test shows this difference to be significant
 at better than the 99\% level. The distributions of V for the BHB and RRL
are shown in Fig.~\ref{fig:Vhist}. The hatched histograms are for the stars
that have negative W-velocities.  Following Shapiro \& Wilks (1965), separate  
 tests of the V-distributions of the BHB and RRL stars with Z $<$ 8 kpc show
that neither have significant differences from normal distributions.

\item 

(b) There is an asymmetry in the ratio of RRL to BHB stars in the sense that the 
 ratio is larger in the sample that has negative W velocities. This effect is 
 significant at better than the 99\% level and suggests that we are not dealing
 with a homeogeneous halo population. 

\item 

(c) The $<$U$>$ space-velocity is significantly negative 
  ($-$72$\pm$21 km s$^{-1}$) for both BHB and RRL stars 
  that have positive W velocities. Those with 
 negative W have the expected zero $<$U$>$.

\end{itemize} 

There is therefore strong evidence that the NGP halo stars with Z$<$ 8 kpc do
 not belong to a homogeneous population. 
It does not seem likely 
that observational selection in the samples could produce the differences
in the kinematic properties between the BHB and RRL stars noted in (b) above. 
The RRL stars cover roughly the same area of the sky as the BHB stars (Fig. 1)
 and the six RRL with positive W are distributed widely in both R.A. and 
Declination. The lack of homogeneity must therefore be present over much of
the survey area. We noted in Sec. 2.3 that stars with R.A. $>$ 13$^{h}$:00$^{m}$
showed predominantly negative radial velocities. There are 6 RRL stars with 
these R.A. and Z $<$  8 kpc. Their heliocentric space-velocities and 
dispersions are:

\begin{eqnarray*}
<U> & = & +010\pm63~~km~s^{-1}~~~\sigma_{u} = 128\pm37~~km~s^{-1} \\
<V> & = & -417\pm74~~km~s^{-1}~~~\sigma_{v} = 152\pm44~~km~s^{-1} \\
<W> & = & -085\pm32~~km~s^{-1}~~~\sigma_{w} = 059\pm17~~km~s^{-1} 
\end{eqnarray*}

This small group thus shows a highly retrograde $<$V$>$ and a negative
 $<$ W $>$, but their corrected velocity dispersions are not unusual.

\begin{figure}
\includegraphics[width=8.2cm]{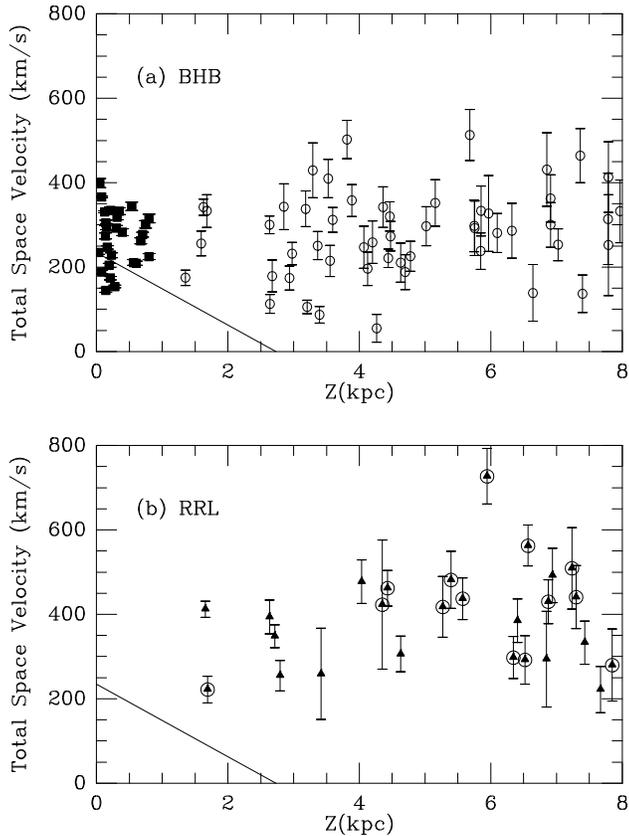}
\caption{Plots of the total space velocity (T) in km~s$^{-1}$ as a
function of the distance from the galactic plane (Z) in kpc for (a) BHB stars 
and (b) RRL stars. The BHB stars at the NGP are shown by open circles and
local BHB stars by filled squares. The RRL stars are shown by filled
triangles; those with W-velocities $<$ $-$50 km~s$^{-1}$ are shown by
encircled triangles. All stars above the line would be classified as
belonging to the HALO2 sample of Martin \& Morrison (1998).
}
\label{f:TSVZ}
\end{figure}

\begin{figure}
\includegraphics[width=8.2cm]{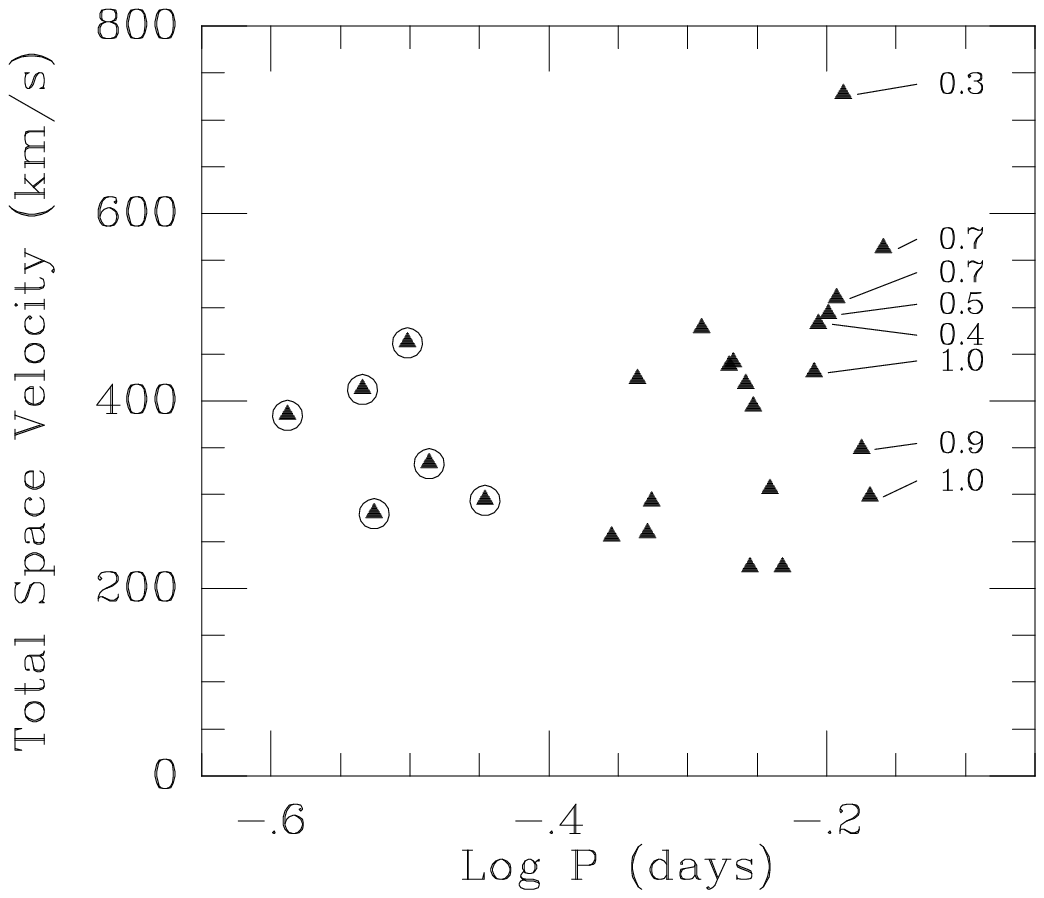}   
\caption{The total space velocity (T) in km~s$^{-1}$ for the RRL  stars
with Z$<$8 kpc plotted against the logarithm of the period P (days). 
The encircled triangles are of type c. A period of 0.60 days
corresponds to log P = $-$0.221. The numbers by the symbols representing the
longer period variables are their visual amplitudes.
}
\label{f:TLGP}
\end{figure}

\subsection{The RRL sample}   

 The 26 RRL stars with Z$<$ 8 kpc have a mean height above the plane of 5.3 kpc.
 Our analysis of the QSO proper motions gave an {\em rms} systematic error 
 in the proper motions of 0.3 mas yr$^{-1}$ in each coordinate.
 This corresponds to a $\sim$2 km~s$^{-1}$ error in V at a distance of 1 kpc. 
Consequently for our Z$<$8 kpc sample, 
 a 2$\sigma$ systematic error in $<$V$>$ from the proper motions should 
 {\em on average} produce 
 a $\sim$20 km~s$^{-1}$ systematic error in V. It does not seem likely therefore
that systematic errors in the proper motions can wholly account for the 
retrograde motion ($\sim$100 km~s$^{-1}$) of the RRL sample. The retrograde V
could also be produced if our RRL distances are 45\% too large. This 
corresponds to a 0.8 mag. error in the modulus of these stars. We used the same
M$_{V}$ for the RRL stars as for the reddest BHB stars\footnote {The 15 BHB
stars that have Z$<$8 kpc and $(B-V)$$>$0.10 have $<$V$>$ = $-$227$\pm$35  and
$\sigma_{v}$ = 116$\pm$21 km~s$^{-1}$ so do not show a retrograde
motion.}.        It does not seem likely, therefore, that our M$_{V}$ can have
such a large  error. Neither does it seem likely that the mean apparent
magnitudes of the RRL stars could have as much systematic error since  they are
based on multiple photoelectric observations.

The total space velocities (T) of our program stars were calculated from the
space velocities U, V \& W  (Tables \ref{t:bhb2} and \ref{t:rrl2})~by: 

\begin{equation}
T =  \sqrt{U^2 + V^2 +W^2}   
\end{equation}

These total space velocities are plotted against the height above the 
galactic plane (Z) for the BHB stars and RRL stars in 
Figs. \ref{f:TSVZ}(a) and \ref{f:TSVZ}(b) respectively. The Local~A  sample of 
nearby BHB stars 
is also plotted as filled squares in Fig. \ref{f:TSVZ}(a). The encircled 
triangles are the RRL stars that have W-velocities $<$ $-$50 km~s$^{-1}$.
The line:

\begin{equation}
T (km s^{-1}) =  235 - 86 \times Z   
\end{equation}

\par\noindent
shows the boundary that is 
used by Martin \& Morrison to separate their HALO2 from their DISK2
populations; all our program stars belong to their HALO2 population.
Most of our stars have values of T that are well below the Galactic escape
velocity of from 500 to 600 km s$^{-1}$ (Smith et al. 2006). Possible 
exceptions are CK COM and SA 57-47 that have T = 563$\pm$49 and 727$\pm$66
km s$^{-1}$ respectively. 
Large total space-motions are associated with (a) stars with 
4$<$Z$<$8 kpc; (b) RRL stars rather than BHB stars; (c) RRL stars
with large negative W-velocities and (d) RRL stars whose periods are
greater than 0.60 days (Fig. \ref{f:TLGP} and Table \ref{t:uvwb}). 
 Among the longer period RRL stars,
 the ones with lower amplitudes have the larger T  (Fig. \ref{f:TLGP}). For 
 periods greater than 0.60 days, we would expect those of lower amplitude to
 belong to Oosterhoff Class I and the higher amplitude variables to belong to
 Oosterhoff Class II. 
 The mean period  $<P_{ab}>$ of the 20 type {\em ab} RRL that have Z $<$ 8 kpc 
  is 0.$^{d}$57. 
 Our assumed RRL absolute magnitudes are linked to  the RRL 
  in the LMC whose $<P_{ab}>$ are
  0.$^{d}$583 and 0.$^{d}$573 from the MACHO (Alcock et al. 1996) and OGLE 
 (Soszy\'{n}ski et al. 2003) surveys respectively. The $<P_{ab}>$ of the RRL
 stars in Galactic globular clusters is 0.$^{d}$585 (Clement et al. 2001);
 65\% of these variables are Oosterhoff type I \footnote{The RRL stars in most
 of the intrinsically faint, 
 metal-poor dSph galaxies have $<P_{ab}>$ $\geq$ 0.$^{d}$60; the Sgr dSph
 on the other hand, has $<P_{ab}>$ = 0.$^{d}$574 (Cseresnjes 2003).}.
 This suggests that the LMC contains predominantly Oosterhoff type I variables.
 This is confirmed by their Fourier types (Alcock et al., 2004) and so our 
 distances should be correct for Oosterhoff I variables which (judging from 
 their $<P_{ab}>$)  probably comprise most of our sample. If, however, the
 Oosterhoff II variables are as much as $\sim$0.2 mag brighter than the 
 Oosterhoff I variables (Lee and Carney 1999) 
then their distances and their total space velocities
will have been underestimated. This may explain some of the dispersion in T
for the RRL stars with P $>$ 0.60 days.

\subsection{The Galactocentric space motions}   

 The Galactocentric space motions of our program stars in the cylindrical
 vectors V$_{R}$, V$_{\phi}$ and V$_{z}$ have been calculated according 
 to the equations given by Kepley et al. (2006) which were chosen to be
 compatible with those used by Helmi et al. (1999). These velocities are
 given in cols. (7), (8) \& (9) respectively of Tables \ref{t:bhb2} and 
 \ref{t:rrl2}. The Z-coordinate in this galactocentric system is in the   
 opposite direction to that in our heliocentric system so that the 
 galactocentric V$_{z}$ has the opposite sign to W in the heliocentric 
 system. Galactocentric V$_{\phi}$ and heliocentric V are respectively negative
 and increasingly negative for {\em retrograde} orbits. The {\em mean} values
 of these space-velocities $<$V$_{R}$$>$, $<$V$_{\phi}$$>$ \& $<$V$_{z}$$>$
 for the program stars with Z$<$ 8 kpc and their corresponding dispersions 
 (after correction as described in Sec. 3.1) are given in Table \ref{t:uvwc}.
 It is seen that the BHB stars show no galactocentric rotation $<$V$_{\phi}$$>$
 = $-$005$\pm$015 km~s$^{-1}$ in agreement with the analysis of the radial
 velocities of 1170 BHB stars from the Sloan Digital Sky Survey (SDSS) by
 Sirko et al. (2004b). Table \ref{t:uvwc} also gives the dispersions in 
 cylindrical coordinates derived by Sirko et al. for three of their samples
 which we call SDSS(a), SDSS(b) and SDSS(c). These refer to their whole sample,
 stars with $g$ $<$ 18 and stars with $g$ $<$ 16 respectively. The dispersions 
 for our BHB are similar to those found for the BHB samples of Sirko et
 al. (2004b) and have $\sigma_{R}$ $\sim$ $\sigma_{\phi}$ whereas non-BHB
 samples of the local halo ellipsoid (Sirko et al. 2004b, Table~1) are like
 our RRL sample and have $\sigma_{R}$ $>$ $\sigma_{\phi}$. There are only
 6 RRL stars in our sample with positive V$_{\phi}$ compared with 20 with 
 negative velocities; the corresponding numbers for the BHB sample are 25 and 
 27. A Fisher's 2$\times$2 test shows that this difference is significant at
 the 96\%-level. A Shapiro-Wilks test of the V$_{\phi}$-distribution of the RRL 
 stars shows no significant departure from normality. {\em It therefore seems
 unlikely that the retrograde $<$V$_{\phi}$$>$ of these stars is caused by
 the inclusion of a few RRL stars  with highly retrograde orbits but rather is
 a property of the whole sample.}

\subsection{Stars with Z $>$ 8 kpc}    

In this paper we have chosen not to analyze the data for the 39 stars with 
Z $>$ 8 kpc. At 10 kpc, the random errors in the proper motions produce errors
 in the space motions that are comparable with the space motions themselves 
 and the estimated 2$\sigma$ systematic error will be $\sim$40 km~s$^{-1}$.
 Currently we do not have radial velocities for all these fainter stars and 
 those whose velocities have errors $\geq$25 km~s$^{-1}$ should be 
 re-observed.

\begin{table*}
\caption{Mean galactocentric space-velocities V$_{R}$, V$_{\phi}$ \&
 V$_{z}$ (km s$^{-1}$) and their corrected dispersions $\sigma_{R}$,
 $\sigma_{\phi}$ \& $\sigma_{z}$  for our program BHB \& RRL stars and for
comparison the SDSS BHB stars of Sirko et al. (2004b) (a) whole sample,
 (b) with $g$ $<$ 18, and (c) with $g$ $<$ 16. 
 }
\label{t:uvwc}
\begin{center}
\leavevmode
\begin{tabular}[h]{cccccccc} 
\noalign{\smallskip}
\hline 
\noalign{\smallskip}
Star  &  n &
  $<$V$_{R}$$>$  &   $\sigma_{R}$ &
  $<$V$_{\phi}$$>$  &   $\sigma_{\phi}$ &
  $<$V$_{z}$$>$  &   $\sigma_{z}$   \\
Type   &     &    &  &  &  &  &   \\  
 (1) & (2) & (3)& (4)& (5) &(6) & (7) &  (8)   \\  
\noalign{\smallskip}
\hline 
\noalign{\smallskip}
      &  &  &  &  &  &  &      \\
 RRL & 26 & $-$11$\pm$35& 162$\pm$23 &$-$095$\pm$29& 122$\pm$ 17  &
$+$043$\pm$17  & 079$\pm$ 11  \\
 BHB & 52 & $-$20$\pm$16& 101$\pm$10 &$-$005$\pm$15&  93$\pm$ 9   &
$+$005$\pm$12  & 081$\pm$ 08  \\
 All & 78 & $-$17$\pm$15& 124$\pm$10 &$-$035$\pm$14& 111$\pm$ 9   &
$+$017$\pm$10  & 081$\pm$ 09 \\
      &  &  &  &  &  &  &      \\
SDSS(a) & 1170 & $\cdots$    & 94$\pm$08 &$\cdots$     & 098$\pm$ 18  &
$\cdots$       & 098$\pm$ 05 \\
SDSS(b) &  773 & $\cdots$    & 97$\pm$10 &$\cdots$     & 109$\pm$ 18  &
$\cdots$       & 101$\pm$ 06 \\
SDSS(c) &  227 & $\cdots$    &109$\pm$20 &$\cdots$     & 122$\pm$ 24  &
$\cdots$       & 094$\pm$ 12 \\

\noalign{\smallskip}
\hline 
\end{tabular}\end{center}\end{table*}

\section{Summary and Conclusions}

 The kinematics of the BHB and RRL stars in the NGP show that this halo
 sample is not homogeneous. The BHB stars show essentially no galactic rotation 
 in agreement with Sirko et al. (2004b) while the RRL  stars show a definite
 retrograde rotation. Our whole sample of 117 stars includes stars
 out to 16 kpc, but we restricted our discussion to the 52 BHB stars and 26 RRL
 stars with Z $<$ 8 kpc whose space motions are the more reliable.
 This sample of 78 stars has a heliocentric $<$V$>$ = $-$268$\pm$14 which is 
 in reasonable agreement with the retrograde rotation of $-$55 km~s$^{-1}$ 
 found by Majewski (1992) and Majewski et al. (1996) for their SA-57 sample of 
 subdwarfs at a similar $<$Z$>$. Both our BHB and RRL sample and their subdwarf 
 sample show qualitatively similar dependencies of U$_{hel}$ on W$_{hel}$
 and an excess of stars with negative W$_{hel}$ (Fig. \ref{f:UVW}). 
 We also find that our BHB stars  with Z$<$ 8 kpc ($<$Z$>$ = 4.7 kpc) have
 $\sigma_{r}$ = 101$\pm$10,  
 $\sigma_{\phi}$ = 93$\pm$09 and 
 $\sigma_{z}$ = 81$\pm$08. These dispersions are similar to those
 of the BHB sample of Sirko et al. (2004b) with $g$ $\leq$ 18 which 
 has $\sigma_{r}$ = 97$\pm$10,  
 $\sigma_{\phi}$ = 109$\pm$18 and 
 $\sigma_{z}$ = 101$\pm$6.These dispersions are more isotropic than local 
 samples of
 halo stars that typically have $\sigma_{r}$$>$$\sigma_{\phi}$$>$$\sigma_{z}$.
 Qualitatively, this is what we might expect from a halo that is more
 flattened towards the plane and more spherical away from the plane and at 
 greater galactocentric distances (Kinman et al. 1965; Wesselink 1987;     
  Preston et al. 1991; Vivas \& Zinn 2006). 
The local BHB sample (Local~A in Table \ref{t:uvwa}) has   
 $\sigma_{r}$ = 129$\pm$18,  
 $\sigma_{\phi}$ = 80$\pm$11 and 
 $\sigma_{z}$ = 101$\pm$14   so that $\sigma_{R}$ $>$ $\sigma_{\phi}$ 
 but (unusually) also has    $\sigma_{z}$ $>$ 
 $\sigma_{\phi}$.  Many of the stars in this sample were 
 selected as early-type high-velocity stars by Stetson (1991) and so Local~A 
 may have some  kinematic bias.

 Our RRL  stars that have positive W-velocities not only have a
 much smaller velocity dispersion ($\sigma_{v}$) than those with negative 
 W-velocity, but the ratio of RRL to BHB stars is significantly lower in this
 group compared with the group that has negative W-velocities.
 This inhomogeneity therefore involves a
 streaming motion that covers a significant part of our survey area and
 which is more pronounced for the stars with an RA $>$ 13$^{h}$:00.
 A similar streaming occurs in the subdwarf sample of Majewski et al. (1996)
 in the same part of the sky.   A model of the Sgr stream 
 (Fig. 1 of Mart\'{\i}nez-Delgardo et al. 2006) also predicts 
 an infall in the Northern Galactic Cap but not quite in the direction of our
 survey and probably refers to more distant stars than most of those in our
 sample. The streaming that we have observed may be connected with the Virgo
 overdensity (Lupton et al. 2005) but this overdensity is not yet 
 sufficiently well defined for this to be more than a speculation.

 The simultaneous use of {\em two} halo tracers (BHB and RRL stars) has shown 
 that the halo has two field star components that seem to parallel the old 
 and young components of the halo globular clusters. This is not at all 
 a new idea. Thus, Wilhelm et al. (1996) (who give references to previous   
 suggestions for two-component halos) 
  analysed the radial velocities of 525 BHB stars 
 and  found an ``inner" flattened halo (Z$<$4 kpc) in prograde (+40~$\pm$17     
 km~s$^{-1}$) rotation and an ``outer" (Z$>$4 kpc) more spherical halo with
 a retrograde (--93$\pm$36 km~s$^{-1}$) rotation . This relatively high 
 retrograde motion for the ``outer" BHB stars does not agree with either our
 results or those of Sirko et al. (2004b). A Z$>$4 kpc sample will be at
 high galactic latitudes where the
 stars lie far from the apices of galactic rotation and their radial velocities
 are relatively insensitive to the galactic rotation vector. This insensitivity
 of the radial velocities and also possible inhomogeneities in the halo may be
 responsible for this discrepancy. On the other hand, the analysis of the 
 space motions of halo stars by Chiba \& Beers (2000) found no evidence for
 a retrograde ``high halo"; possibly because their sample did not
 extend far enough from the solar neighbourhood. 

  The two halo components that fit our data
 are  similar to those discussed by Borkova and Marsakov (2003)
  from their analysis of the space-motions of nearby RRL stars. 
 One component is associated with a strong blue horizontal branch, has no 
 galactic rotation and has  a relatively isotropic velocity dispersion.
 The other component is characterized by having RRL stars whose  $<P_{ab}>$
 shows that they are mostly of Oosterhoff type I. This second component has, 
 on average, a retrograde rotation and a less isotropic velocity 
 dispersion than the first component.  Our  second component also has 
 streaming in its W space-motion which suggests that it consists  of tidal 
 debris. This does not mean that the first component contains no
 RRL stars or the second has no BHB stars but that the BHB stars predominate
 in the first component and the RRL stars in the second. We would expect that
 the first component, having a blue HB, would be associated with Oosterhoff
 type II RRL stars. We must be careful about this assertion, however, because
 there is considerable scatter in the HB-type {\em vs.} $<P_{ab}>$ plot; e.g.
  M~62 has $<P_{ab}>$ = 0.$^{d}$548 (Oosterhoff type I) and yet has a blue HB
  (Clement et al. 2001; Contreras et al. 2005). 

 It would be very desirable to be able to compare all the properties ([Fe/H], 
 space densities, HB type, Oosterhoff type etc) of the halo stars in the North
 Polar Cap with those in the solar neighbourhhod. This is beyond the scope
 of this paper and would require considerable attention to selection effects.
 As an example, all RRL  surveys miss variables with amplitudes below a 
 certain limit. Our photometry of the BHB stars showed that a number of these
 were low-amplitude RRL  stars so, in this respect, our polar sample is 
 unusually complete and contains low-amplitude long-period type {\em ab}
  and type {\em c} variables that were missed in earlier surveys. 
 Our results do not necessarily conflict with those of earlier surveys that
 have sampled larger volumes of space.
 The volume of space that we have sampled is relatively small. At the mean 
 distance of 5 kpc, the width of the field is no more than $\sim$1 kpc. This
 may account for the pronounced lack of 
 uniformity of the halo that we have found at the NGP and makes 
 it essential that similar surveys should be made at other key locations
 such as the South Galactic Pole and the Anticenter and if possible that they
 should include more than one halo tracer.

\section*{Acknowledgments}

We thank D.R. Soderblom for kindly making  available the 
program to calculate the UVW space motions. We are also grateful to Dr C.
Clement for helpful comments on fundamentalizing RRL periods and 
A. Sollima for assistance in providing us with the infrared ZAHBs. 
This research has made use of 2MASS data provided by the NASA/IPAC 
Infrared Science Archive, which is operated by the Jet Propulsion Laboratory, 
California Institute of Technology, under contract with the National 
Aeronautics and Space Administration.\\
The GSC-II is a joint project of the Space Telescope Science
Institute (STScI) and the INAF-Osservatorio Astronomico di Torino
(INAF-OATo). We acknowledge the GSC-II team and in particular M.G.
Lattanzi, B. McLean and R.L. Smart for their valuable support to this study.\\
This work is partly based on observations made with the Italian Telescopio 
 Nazionale Galileo (TNG) operated on the island of La Palma by the Fundacion 
Galileo Galilei of the INAF (Istituto Nazionale di Astrofisica) at the Spanish
Observatorio del Roque de los Muchachos of the Instituto de Astrofisica de
Canarias.\\
This work has been partly supported by the MIUR (Mi\-ni\-ste\-ro 
dell'Istruzione, dell'Universit\`a e della Ricerca) under 
PRIN-2001-1028897 and by PRIN-INAF 2005 1.06.08.02 and 03.

%%%%%%%%%%%%%%%% Big  Tables %%%%%%%%%%%%%%%%%%%%%%%%%%%%%%%%%%%%%%%%%%

\newpage

\begin{center}
\begin{landscape} 

\topmargin 40mm
\thispagestyle{empty}

\begin{table*} 
\caption{The BHB stars of our sample. The coordinates are for J2000. 
The parallaxes $\Pi_{HBP}$, $\Pi_{HBC}$ and $\Pi_{HBK}$ are defined 
in Sect. 2.2.3; the error $\sigma_{\Pi}$ refers to the adopted parallax 
$\Pi_{HBP}$. Proper motions and their errors (columns 12-15) are in 
$mas\,yr^{-1}$. Radial velocities and their errors (columns 16 and 17) are 
in km s$^{-1}$.  
}
\label{t:bhb1}
\begin{tabular}{@{}rcccccccccccccccc@{}} 
\hline 
ID  & RA&DEC&V & B-V & E(B-V) & K  & $\Pi_{HBP}$ & $\sigma_{\Pi}$&$\Pi_{HBC}$&$\Pi_{HVK}$&$\mu_{\alpha}$&$\sigma \mu_{\alpha}$&
$\mu_{\delta}$&$\sigma \mu_{\delta}$ &RV & $\sigma$RV \\
(1) &(2)&(3) &(4) &(5) &(6) &(7) &(8) &(9) &(10) &(11) &(12) &(13) &(14) &(15)& (16)& (17)  \\
\hline 

    16549-51 &  12 01 02.4 &34 37 40  &13.67 &   0.078 & 0.016 &  13.30 A & 0.271&0.014&0.280&0.258& --15.8&  1.9 & --7.9 & 1.1&   46  &  50\\
      AF-003 &  12 03 20.1 &32 01 38  &16.48 &   0.230 & 0.019 &  15.35 C & 0.066&0.003&0.067&0.067&   1.0&  4.1 & --3.8 & 4.6&  --75  &  40\\ 
      AF-006 &  12 04 43.1 &29 41 50  &13.64 &   0.007 & 0.016 &  13.54 A & 0.309&0.026&0.332&0.282& --9.2&  3.5& --20.0 & 1.5&   28  &  10\\ 
      AF-011 &  12 06 19.5 &31 51 37  &15.85 &   0.101 & 0.019 &  14.72 U & 0.098&0.004&0.100&0.090&   2.4&  3.4& --13.7 & 6.2&  --92  &  10\\ 
      AF-013 &  12 07 11.2 &28 58 52  &16.78 &   0.115 & 0.019 &  17.18 U & 0.063&0.003&0.064&0.086& --3.8&  2.2 &--4.0  & 1.5&  128  &  40\\       
      AF-727 &  12 08 42.0 &36 02 08  &13.93 &   0.137 & 0.019 &  13.52 A & 0.229&0.009&0.231&0.228&   1.4&  1.6 &--2.0& 1.8&  --24  &  10\\ 
      AF-022 &  12 10 08.2 &30 53 00  &16.37 &   0.082 & 0.019 &  15.96 U & 0.078&0.004&0.081&0.074& --1.9&  2.1 &--5.9  & 1.0&  --40  &  10\\ 
      AF-729 &  12 10 26.4 &34 32 54  &15.24 &   0.146 & 0.013 &  14.70 A & 0.123&0.005&0.123&0.121& --3.6&  1.9 &--7.8  & 1.9&  --32   & 40\\ 
      AF-029 &  12 12 59.8 &33 43 13  &14.85 &   0.131 & 0.013 &  14.22 A & 0.148&0.006&0.150&0.144&   0.4&  2.8 &--3.9  & 1.7&  --62  &  10\\       
      AF-030 &  12 13 01.9 &29 09 05  &14.85 &   0.044 & 0.023 &  14.48 A & 0.169&0.012&0.178&0.152& --7.8&  2.4 &--8.9  & 1.8&   42  &  10\\ 
    16022-26 &  12 14 29.2 &27 23 52  &14.65 &   0.097 & 0.023 &  14.18 A & 0.172&0.008&0.177&0.164& --4.8&  1.6 &--9.4  & 2.5&  --29  &  50\\ 
      AF-038 &  12 15 01.6 &32 02 21  &15.63 &   0.049 & 0.016 &  15.23 B & 0.114&0.007&0.120&0.104& --6.8&  1.7 &--7.9  & 3.3&  --78  &  10\\ 
      AF-039 &  12 15 04.8 &30 06 18  &16.65 & --0.006 & 0.019 &  15.66 D & 0.081&0.007&0.087&0.062& --5.9&  1.5 &--1.0 & 1.6&	58  &  40\\ 
      AF-041 &  12 15 35.4 &32 35 38  &15.02 &   0.168 & 0.013 &  14.26 A & 0.134&0.005&0.134&0.132& --9.2&  1.6 &--9.0 & 1.9&   95  &  10\\       
      AF-045 &  12 16 49.9 &28 56 06  &13.45 &   0.129 & 0.026 &  12.86 A & 0.292&0.012&0.296&0.281& --4.9&  1.2& --2.0  & 1.3&   17  &  10\\ 
      AF-046 &  12 16 53.7 &30 04 09  &15.98 &   0.130 & 0.023 &  15.69 D & 0.090&0.004&0.091&0.092& --3.9&  1.6 & --2.0 & 3.4&   64  &  10\\ 
      AF-048 &  12 17 20.9 &32 11 27  &13.92 & --0.008 & 0.013 &  13.85 A & 0.278&0.025&0.301&0.248& --7.8&  2.1 & --9.9 & 1.8&   22  &  10\\ 
      AF-050 &  12 17 34.8 &29 19 24  &16.57 &   0.222 & 0.026 &  15.89 U & 0.065&0.003&0.065&0.066& --2.0&  3.2 &--4.9  & 2.2&   27  &  40\\       
      AF-052 &  12 18 05.5 &30 17 25  &14.74 &   0.081 & 0.019 &  14.30 A & 0.166&0.009&0.172&0.157& --8.9&  4.2 &--6.9  & 2.7&  --66  &  10\\ 
      AF-053 &  12 18 08.5 &33 24 37  &15.19 &   0.026 & 0.013 &  15.09 B & 0.144&0.011&0.153&0.137& --1.8&  1.3 &--13.0  & 2.9&   11  &  10\\ 
      AF-063 &  12 22 34.2 &30 08 36  &16.29 &   0.092 & 0.019 &  16.19 U & 0.080&0.004&0.083&0.083& --4.0&  1.8 & --3.9  & 1.0& --103  &  10\\ 
    16026-28 &  12 23 02.6 &27 27 16  &13.65 &   0.009 & 0.019 &  13.50 A & 0.310&0.026&0.333&0.278& --0.5&  1.7 &--4.2  & 1.3&  --85  &  10\\ 
      AF-754 &  12 23 24.9 &36 31 09  &15.25 &   0.106 & 0.013 &  14.84 A & 0.126&0.005&0.128&0.123& --3.0&  1.3 &--6.0  & 3.6&  --15  &  40\\ 
      AF-755 &  12 23 26.9 &36 41 12  &15.46 &   0.034 & 0.013 &  15.06 B & 0.126&0.009&0.133&0.112& --4.9&  2.5 &--6.0  & 3.1& --120  &  40\\ 
      AF-068 &  12 26 10.3 &30 07 13  &14.63 &   0.108 & 0.019 &  14.12 A & 0.170&0.008&0.173&0.163& --4.0&  2.6& --5.8  & 1.5& --134  &  10\\       
      AF-070 &  12 26 37.4 &28 49 24  &15.19 &   0.047 & 0.023 &  14.98 B & 0.144&0.010&0.151&0.135& --5.5&  1.4& --4.8  & 2.1&  181  &  10\\ 
      AF-073 &  12 28 56.2 &32 59 29  &16.81 &   0.158 & 0.016 &  15.98 D & 0.059&0.002&0.060&0.058&   1.0&  1.6 &--4.0  & 1.5&  --41  &  40\\ 
      AF-075 &  12 30 33.4 &30 59 59  &15.36 &   0.115 & 0.016 &  14.89 A & 0.120&0.005&0.122&0.116&   0.3&  2.8 &--10.9 & 2.0&  --50  &  10\\ 
      AF-076 &  12 30 44.4 &30 29 29  &15.72 &   0.128 & 0.019 &  15.03 B & 0.101&0.004&0.102&0.097& --1.8&  2.5 &--6.0  & 0.7&   37  &  10\\       
      AF-077 &  12 30 51.9 &32 34 07  &15.60 &   0.102 & 0.013 &  15.10 B & 0.108&0.005&0.110&0.103& --0.6&  3.1& --8.2  & 4.0& --159  &  10\\ 
      AF-078 &  12 31 10.6 &30 51 26  &14.05 &   0.092 & 0.016 &  13.51 A & 0.224&0.011&0.230&0.211&   2.1&  0.9 &--5.9  & 2.0& --178  &  10\\ 
      AF-769 &  12 34 07.2 &33 57 01  &13.66 &   0.062 & 0.019 &  13.32 A & 0.281&0.017&0.293&0.262&--17.8&  2.0&--16.6  & 2.5&    9  &  10 \\ 
    16026-67 &  12 36 21.8 &27 16 34  &13.40 &   0.006 & 0.019 &  13.28 A & 0.350&0.030&0.377&0.314&--15.7&  2.3&--19.8  & 3.8&  --29  &  10\\
      AF-100 &  12 39 18.2 &29 37 53  &13.74 &   0.090 & 0.013 &  13.35 A & 0.257&0.012&0.263&0.247&--16.9&  1.5 &--9.1  & 2.2&   57  &  10\\
    16466-08 &  12 40 41.7 &33 52 27  &15.22 & --0.014 & 0.016 &  15.09 C & 0.157&0.015&0.170&0.135& --7.9&  2.8 &--5.2  & 1.4&   17  &  40\\ 
      AF-108 &  12 41 17.6 &29 09 15  &13.86 &   0.035 & 0.013 &  13.67 A & 0.262&0.018&0.277&0.245&--17.9&  1.8&--20.9  & 1.8&  --70  &  10\\ 
      AF-111 &  12 41 38.4 &33 20 55  &14.33 &   0.087 & 0.016 &  14.08 A & 0.198&0.010&0.203&0.195&--12.0&  1.9 & 3.2  & 1.9&  --18  &  10\\ 
      AF-112 &  12 42 07.9 &30 55 41  &13.21 &   0.006 & 0.016 &  13.09 A & 0.378&0.032&0.406&0.340&   1.0&  1.9 &--9.0  & 1.6&    2  &  10\\       
      AF-113 &  12 42 12.4 &31 56 48  &11.78 &   0.139 & 0.016 &  11.22 A & 0.611&0.025&0.614&0.598&--25.9&  1.9&--35.8 & 1.7&	 4  &  10\\ 
      AF-115 &  12 42 39.3 &33 23 52  &15.42 &   0.068 & 0.016 &  15.02 B & 0.123&0.007&0.127&0.115& --5.8&  1.7 &--8.6  & 1.2&   13  &  10\\
    16466-15 &  12 43 41.8 &32 21 28  &13.14 &   0.089 & 0.013 &  12.71 A & 0.339&0.016&0.348&0.324& --5.9&  2.5 &--3.8  & 4.1&  145  &  10\\ 
      AF-130 &  12 44 18.6 &29 26 20  &15.92 &   0.180 & 0.013 &  15.25 B & 0.088&0.004&0.088&0.087& --1.0&  1.9 &--4.0  & 0.9&  106  &  10\\ 
      AF-131 &  12 44 18.6 &30 55 22  &15.58 &   0.119 & 0.016 &  15.01 B & 0.108&0.005&0.109&0.104& --4.0&  1.9 &--6.0  & 1.7&   47  &  10\\ 
      AF-134 &  12 44 42.2 &32 29 55  &12.76 &   0.170 & 0.013 &  12.04 A & 0.378&0.015&0.378&0.373&   6.2&  1.9&--23.1  & 1.1&  --27  &  10\\ 
\hline
\end{tabular}
\end{table*}
\end{landscape}

%\newpage 

\thispagestyle{empty}
\topmargin 40mm

\addtocounter{table}{-1} 
\begin{landscape}
\begin{table*}
\caption{continued }
\label{t:bhb1}
\begin{tabular}{@{}rcccccccccccccccc@{}} 
\hline 
ID  & RA&DEC&V & B-V & E(B-V) & K  & $\Pi_{HBP}$ & $\sigma_{\Pi}$&$\Pi_{HBC}$&$\Pi_{HVK}$&$\mu_{\alpha}$&$\sigma \mu_{\alpha}$&
$\mu_{\delta}$&$\sigma \mu_{\delta}$ &RV & $\sigma$RV \\
(1) &(2)&(3) &(4) &(5) &(6) &(7) &(8) &(9) &(10) &(11) &(12) &(13) &(14) &(15)& (16)& (17)  \\
\hline 
    16031-44 &  12 46 15.6 &27 10 47  &13.80 &   0.145 & 0.013 &  13.23 A & 0.238&0.010&0.239&0.234&   1.4&  1.2&--13.0  & 3.2&   11  &  10\\ 
    15622-48 &  12 46 43.7 &27 17 17  &14.47 &   0.153 & 0.013 &  13.97 A & 0.174&0.007&0.174&0.174& --3.1&  1.4&--10.2  & 2.6&  --69  &  10\\ 
    15622-07 &  12 46 47.3 &27 27 44  &12.86 &   0.121 & 0.013 &  12.44 A & 0.374&0.016&0.378&0.369& --5.1&  2.3&--12.9  & 3.3&  --37  &  10\\ 
      AF-138 &  12 47 37.2 &33 07 28  &14.72 & --0.009 & 0.013 &  14.65 A & 0.193&0.017&0.209&0.171& --8.9&  3.0 &--10.5 & 1.6&  101  &  10\\ 
    15622-09 &  12 48 58.8 &26 48 04  &13.08 &   0.122 & 0.010 &  12.48 A & 0.336&0.014&0.339&0.324& --1.7&  1.4&--14.0  & 2.5& --121  &  10\\ 
      AF-797 &  12 50 23.2 &36 20 53  &15.42 &   0.074 & 0.016 &  15.07 A & 0.122&0.006&0.126&0.116& --8.0&  2.2 &--6.0  & 0.3&   85  &  40\\ 
      AF-804 &  12 51 52.7 &36 40 42  &14.07 &   0.098 & 0.016 &  13.64 A & 0.221&0.010&0.225&0.212&--13.0&  2.3& --5.9  & 1.0&   94  &  10\\       
    SA57-001 &  12 52 18.7 &28 19 58  &14.43 &   0.155 & 0.010 &  13.76 A & 0.176&0.007&0.176&0.173& --1.9&  2.8&--18.4  & 1.8& --124  &  10\\ 
    SA57-006 &  12 53 57.3 &29 38 08  &13.99 &   0.083 & 0.010 &  13.58 A & 0.229&0.011&0.235&0.218&--14.6&  2.7& --5.6  & 2.2&  112  &  10\\ 
    SA57-007 &  12 54 03.7 &28 51 47  &12.18 & --0.010 & 0.013 &  12.11 A & 0.626&0.058&0.678&0.553&--13.9&  1.6& --28.5 & 2.2&  --92  &  50\\ 
    SA57-009 &  12 54 18.0 &28 43 42  &15.96 &   0.204 & 0.013 &  15.14 B & 0.085&0.003&0.085&0.085& --9.6&  2.7 &--9.1  & 1.7&   47  &  10\\ 
    SA57-017 &  12 56 23.8 &27 28 39  &13.41 &   0.084 & 0.016 &  13.00 A & 0.304&0.015&0.313&0.289&--25.4&  4.7& --10.6 & 3.5&   --4  &  10\\
    SA57-021 &  12 57 34.6 &31 48 45  &16.62 & --0.066 & 0.016 &  16.43 U & 0.097&0.011&0.107&0.069& --6.0&  2.2 & 5.8  & 3.3&  --70  &  40\\
    SA57-029 &  12 59 13.6 &30 36 19  &16.31 &   0.085 & 0.010 &  15.72 D & 0.079&0.004&0.081&0.073& --2.9&  1.6 &--4.0  & 1.8& --182  &  40\\
    SA57-032 &  12 59 43.3 &31 20 30  &15.13 &   0.155 & 0.013 &  14.63 A & 0.128&0.005&0.128&0.128& --7.8&  1.5 &--8.0  & 2.5&  --14  &  10\\ 
    SA57-036 &  13 00 41.2 &28 52 37  &16.23 &   0.062 & 0.010 &  15.37 B & 0.084&0.005&0.087&0.075& --3.9&  1.9 &  --5.9 &2.7&   --5  &  10\\        
    SA57-040 &  13 03 13.6 &28 40 11  &15.14 &   0.080 & 0.010 &  14.67 A & 0.135&0.007&0.139&0.127& --2.9&  1.5& --2.5  & 1.2&   30  &  10\\    
    SA57-041 &  13 03 20.6 &28 43 28  &16.33 &   0.092 & 0.010 &  15.62 C & 0.077&0.004&0.079&0.072& --4.9&  1.9 &--3.5  & 1.8&  127  &  10\\ 
    SA57-045 &  13 04 22.7 &31 30 25  &15.31 &   0.108 & 0.010 &  15.12 B & 0.122&0.005&0.123&0.125&   1.0&  1.4 & --7.9 & 1.0& --135  &  10\\     
      AF-825 &  13.04 56.3 &35 03 31  &14.32 &   0.051 & 0.013 &  14.08 A & 0.207&0.013&0.216&0.196& --5.9&  0.8 &--5.1  & 1.4& --139  &  50 \\
    SA57-046 &  13 05 08.7 &27 48 01  &16.71 &   0.148 & 0.013 &  16.76 U & 0.062&0.002&0.062&0.072& --4.9&  1.7 &--4.4  & 2.2&  151  &  40\\   
    SA57-055 &  13 07 19.4 &30 25 56  &16.03 &   0.140 & 0.013 &  15.58 C & 0.086&0.003&0.086&0.085& --1.0&  3.1 &--5.9  & 0.7& --224  &  10\\ 
      AF-841 &  13 08 30.3 &32 43 40  &14.06 &   0.091 & 0.013 &  13.62 A & 0.222&0.010&0.227&0.212& --7.8&  1.2&--10.1  & 1.7&  --25  &  10\\ 
      AF-848 &  13 10 48.5 &32 28 19  &11.75 &   0.001 & 0.010 &  11.64 A & 0.732&0.061&0.786&0.661& --9.8&  3.0&--18.0  & 2.8& --116  &  10\\
    SA57-066 &  13 11 29.4 &31 47 06  &14.68 &   0.116 & 0.013 &  14.08 A & 0.163&0.007&0.165&0.156& --3.0&  3.4 &--3.9  & 2.1& --243  &  10\\
      AF-854 &  13 12 01.6 &37 26 55  &15.11 &   0.058 & 0.013 &  14.77 A & 0.142&0.008&0.148&0.133&--10.9&  1.8 &--0.0  & 1.4&  --10  &  50\\
    SA57-080 &  13 13 10.2 &30 45 00  &14.16 &   0.077 & 0.013 &  13.90 A & 0.215&0.011&0.222&0.210& --8.0&  1.2 &--4.8  & 2.5&  --46  &  10\\ 
      AF-866 &  13 13 39.3 &34 27 18  &13.93 &   0.059 & 0.010 &  13.64 A & 0.243&0.014&0.252&0.230& --6.0&  1.8& --11.0 & 2.8& --41  &  50\\
    SA57-087 &  13 14 29.3 &31 42 15  &15.97 &   0.093 & 0.010 &  15.45 C & 0.091&0.004&0.093&0.086& --4.5&  2.6 &--3.8  & 2.6& --112  &  10\\    
      AF-900 &  13 20 25.1 &33 57 56  &13.71 &   0.013 & 0.013 &  13.55 A & 0.293&0.023&0.313&0.265& --9.6&  1.5&--10.8  & 0.6&   92  &  50\\
    SA57-111 &  13 20 37.2 &30 12 14  &16.47 &   0.094 & 0.016 &  15.66 U & 0.073&0.003&0.075&0.068& --5.9&  1.3& --4.0  & 1.5& --138  &  10\\
      AF-909 &  13 22 12.0 &31 21 31  &15.26 &   0.023 & 0.013 &  15.09 A & 0.141&0.010&0.149&0.130& --1.9&  1.3 &--5.9  & 1.2& --144  &  10\\
      AF-914 &  13 23 21.0 &37 47 05  &12.27 & --0.003 & 0.010 &  12.16 A & 0.581&0.050&0.626&0.519&--36.2&  3.4&--15.0  & 3.4&   95  &  50\\
      AF-916 &  13 24 15.4 &33 22 13  &14.00 &   0.181 & 0.010 &  13.69 A & 0.210&0.008&0.211&0.222& --0.2&  1.3 & --6.0 & 1.7& --132  &  50\\
      AF-918 &  13 25 02.6 &29 46 18  &14.08 &   0.028 & 0.013 &  13.87 A & 0.240&0.017&0.254&0.220& --7.9&  2.3& --5.9  & 1.6&  --24  &  50 \\ 
\hline
\end{tabular}           
\end{table*}
\end{landscape}

%----------------------------------------------------------- 

\newpage 

\thispagestyle{empty}
%\topmargin 40mm

%\centering
%\begin{center}

\begin{landscape}
\begin{table*}
\caption{The RRL stars of our sample. $<$K$>$ magnitudes are corrected for 
phase effect as described in the text. The parallaxes labelled $\Pi_{V}$, 
$\Pi_{K}$ and $\Pi_{HB}$ were estimated as described in Sect. 2.2.2. 
The error $\sigma_{\Pi}$ refers to the adopted parallax $\Pi_{V}$. RV and 
$\sigma$RV are in km s$^{-1}$. 
}
\label{t:rrl1}
\begin{tabular}{@{}rccccccccccccccccc@{}} 
\hline 
ID  & RA & DEC & logP & [Fe/H] & $<$V$>$ & $<$K$>$ & E(B-V) &$\Pi_{V}$ &$\sigma_{\Pi}$ &$\Pi_{K}$ &$\Pi_{HB}$&$\mu_{\alpha}$&$\sigma \mu_{\alpha}$&$\mu_{\delta}$ 
& $\sigma \mu_{\delta}$ & RV & $\sigma$RV \\
(1) &(2)&(3) &(4) &(5) &(6) &(7) &(8) &(9) &(10) &(11) &(12) &(13) &(14) &(15) &(16)&(17)&(18)\\
\hline 
 GR-Com & 12 05 18.6  &  27 56 58   &--0.282 & ----- & 16.33 & 15.14 B& 0.021 & 0.071 &0.004& 0.075 & 0.072& --3.8&  1.3&  --6.0&   1.7&   -----& ----- \\ 
 GH-Com & 12 05 24.5  &  27 38 22   &--0.352 & ----- & 15.38 & 14.42 A& 0.020 & 0.110 &0.007& 0.114 & 0.112& --1.7&  1.5&  --8.9&   2.3&   -----& ----- \\ 
 IQ-Com & 12 06 03.6  &  27 59 18   &--0.294 & ----- & 15.70 & 14.68 A& 0.020 & 0.095 &0.006& 0.095 & 0.095& --3.9&  1.8&  --1.1&   1.1&   -----& ----- \\ 
NSV5476 & 12 09 17.0  &  33 39 36   &--0.486$^{\ast}$ & ----- & 14.96 & 14.16 A& 0.013 & 0.132 &0.008& 0.128 & 0.135& --4.1&  1.9&  --8.3&   1.1&  --27  &  50 \\
  V-Com & 12 10 15.9  &  27 25 54   &--0.329 &--1.80 & 13.25 & 12.39 A& 0.022 & 0.289 &0.009& 0.280 & 0.296&--15.7&  1.9&  --0.8&  10.6&   +23  &  28    \\
 CD-Com & 12 12 34.0  &  30 48 03   &--0.229 &--1.94 & 15.40 & 14.12 A& 0.018 & 0.105 &0.004& 0.105 & 0.113&--12.7&  1.4& --11.0&   1.9& --203  &  10  \\ 
 AF-031 & 12 13 04.4  &  30 16 38   &--0.519$^{\ast}$ & ----- & 15.49 & 14.74 A& 0.019 & 0.104 &0.006& 0.103& 0.107& --7.8&  0.8 & --7.0&   3.0 &--167 & 50  \\ 
 TU-Com & 12 13 46.9  &  30 59 08   &--0.336 &--1.59 & 13.80  & 13.08 A& 0.018 & 0.227 &0.007& 0.206& 0.233& --0.8&  1.9& --19.7&   9.1&  --98  &  10  \\ 
 CK-Com & 12 14 50.6  &  33 06 07   &--0.159 &--1.85 & 14.62  & 13.50 A& 0.012 & 0.150 &0.008& 0.139& 0.153& --7.4&  1.4& --16.0&   1.2&  --88  &  10   \\ 
 AF-042 & 12 15 39.3  &  29 30 43   &--0.502$^{\ast}$ & ----- & 16.35 & 15.31 U& 0.021 & 0.070 &0.004&0.074& 0.072& --3.9&  2.3&  --4.0&   2.0& -----& -----\\ 
 CL-Com & 12 17 16.3  &  30 58 39   &--0.255 &--1.17 & 15.10 & 13.79 A& 0.016 & 0.129 &0.005& 0.138 & 0.131& --1.0&  2.6&  --5.9&   0.6&  --34  &  10  \\ 
 AT-CVn & 12 18 17.1  &  33 39 56   &--0.446$^{\ast}$ & ----- & 14.76 & 13.76 A& 0.013 & 0.144 &0.009& 0.148 & 0.147& --5.9&  2.1&  --6.1&  4.0&  +94 &  50  \\
 GY-Com & 12 19 26.0  &  28 27 28   &--0.272 & ----- & 15.89 & 14.58 A& 0.026 & 0.087 &0.006& 0.097 & 0.089&  +2.9&  1.5&  --5.9&   1.8&   -----& ----- \\
 GS-Com & 12 24 56.1  &  27 59 08   &--0.276 & ----- & 16.44 & 15.04 A& 0.021 & 0.067 &0.004& 0.078 & 0.069& --1.0&  1.4&  --4.9&   4.7&   -----& ----- \\ 
 RR-CVn & 12 29 07.6  &  34 38 51   &--0.253 &--1.50 & 12.72 & 11.61 A& 0.016 & 0.375 &0.014& 0.372 & 0.381&--19.1&  2.2& --24.7&   3.2&   --5  &  10   \\
  S-Com & 12 32 45.7  &  27 01 46   &--0.232 &--1.79 & 11.64 & 10.60 A& 0.019 & 0.589 &0.014& 0.571 & 0.604&--19.6&  4.4& --18.4&   3.9 & --55  &   4  \\ 
 SV-CVn & 12 35 56.0   & 37 12 26   &--0.175 &--2.20 & 12.65 & 11.46 A& 0.018 & 0.362 &0.015& 0.358 & 0.373& --6.6&  0.9& --25.7&   2.2&   +29  &  28  \\ 
 FV-Com & 12 37 57.2  &  29 58 06   &--0.326 & ----- & 14.65 & 13.46 A& 0.018 & 0.153 &0.010& 0.171 & 0.156&  +2.1&  1.9&  --7.0&   2.0& --186  &  50  \\ 
  U-Com & 12 40 03.4  &  27 29 57   &--0.534$^{\ast}$ &--1.25 & 11.73 & 10.94 A& 0.016 & 0.602 &0.014&0.603  & 0.614&--49.7&  2.7& --16.4&  2.1& --22  &  3  \\ 
 SW-CVn & 12 40 55.1  &  37 05 07   &--0.355 &--1.53 & 12.84 & 11.98 A& 0.013 & 0.352 &0.015& 0.351 & 0.359&--17.8&  2.7&  --6.4&   2.6&  --18  &  21   \\
 DV-Com & 12 43 54.4  &  28 01 16   &--0.267 &--1.74 & 14.84 & 13.63 A& 0.014 & 0.137 &0.003& 0.148 & 0.140&  +1.9&  1.4& --12.0&   2.8& --136  &  10   \\
 AF-791 & 12 47 16.3   & 35 12 06   &--0.209 &--1.68 & 14.77 & 13.55 A& 0.015 & 0.144 &0.009& 0.144 & 0.147&--10.0 & 1.8&   +5.9&   1.6& --200  &  10  \\
 AW-Com & 12 49 18.0  &  28 49 27   &--0.347 & ----- & 15.64 & 14.87 B& 0.015 & 0.095 &0.006& 0.092 & 0.098&  +2.4&  1.0&  --5.0&   2.1&  ----- & -----\\
 TX-Com & 12 50 00.7  &  31 08 25   &--0.270 & ----- & 14.29 & 13.25 A& 0.012 & 0.179 &0.011& 0.177 & 0.182&  +0.2&  2.1& --15.9&   1.4& --118  &  50   \\
 AP-CVn & 12 51 10.6  &  32 58 19   &--0.241 & ----- & 13.90 & 12.70 A& 0.013 & 0.215 &0.014& 0.221 & 0.219&--13.0&  1.5&  --4.9&   2.0&  --20  &  50  \\
 EM-Com & 12 51 38.3  &  30 31 04   &--0.266 &--1.80 & 15.41 & 14.26 A& 0.014 & 0.105 &0.004& 0.110 & 0.107& --1.2&  2.5&  --7.9&   1.6& --127  &  10  \\
 AF-155 & 12 53 51.2  &  32 09 56   &--0.526$^{\ast}$ &--1.56 & 15.05 & 14.22 A& 0.014 & 0.127 &0.008&0.131  & 0.129& --3.4& 2.1&  --5.9&  2.5& --119  & 32  \\
 TY-CVn & 12 54 21.6  &  32 14 34   &--0.290 &--1.23 & 13.60 & 12.65 A& 0.015 & 0.247 &0.016& 0.239 & 0.252&--24.4&  1.6&  --3.3&   3.3&   +74  &  16  \\
 IP-Com & 12 56 30.8  &  29 53 36   &--0.193 &--1.48 & 14.85 & 13.49 A& 0.012 & 0.138 &0.009& 0.146 & 0.141& --3.7&  2.3& --13.8&   3.0& --137  &  30  \\
SA57-19 & 12 56 51.2  &  28 10 35   &--0.588$^{\ast}$ & ----- & 14.58 & 13.88 A& 0.010 & 0.156 &0.009& 0.164 & 0.161& --9.7&  1.6&  --8.0&  1.6&  +49 & 50  \\
 EO-Com & 12 57 22.1  &  28 53 20   &--0.199 &--1.67 & 14.74 & 13.56 A& 0.011 & 0.144 &0.005& 0.142 & 0.147&--11.7&  2.7&  --9.1&   1.4&   +72  &  10  \\
 UV-Com & 12 59 34.1  &  30 19 25   &--0.211 &--1.61 & 15.58 & 14.31 A& 0.011 & 0.099 &0.003& 0.102 & 0.101& --8.8&  1.0&  --2.1&   2.0&   +58  &  10  \\
 TZ-CVn & 13 01 29.2  &  32 05 13   &--0.258 & ----- & 14.17 & 13.32 A& 0.012 & 0.189 &0.012& 0.170 & 0.193& --2.4&  3.1& --15.1&   2.7& --167  &  50  \\
SA57-47 & 13 05 14.5   & 28 37 14   &--0.188 &--1.51 & 14.43 & 13.16 A& 0.011 & 0.168 &0.010& 0.169 & 0.171& --9.8&  1.9& --23.1&   1.9& --166  &  37  \\
SA57-60 & 13 09 29.7  &  27 01 00   &--0.206 & ----- & 14.23 & 12.88 A& 0.017 & 0.185 &0.011& 0.196 & 0.189&--11.0&  2.1& --14.7&   2.6& --109  &  50  \\
 EW-Com & 13 13 01.7  &  31 01 22   &--0.274 &--1.92 & 15.02 & 14.26 A& 0.012 & 0.124 &0.003& 0.111 & 0.128& --8.4&  1.3&  --1.8&   2.2&   +19  &  10  \\
 IS-Com & 13 14 38.8  &  27 56 30   &--0.502$^{\ast}$ & ----- & 13.80 & 12.96 A& 0.013 & 0.225 &0.014& 0.228 & 0.230&--13.4& 1.4& --15.6&  1.8& --163  &  10  \\
 AF-882 & 13 17 03.5  &  36 06 58   &--0.169 & ----- & 14.60 & 13.50 A& 0.011 & 0.155 &0.009& 0.142 & 0.158& --4.1&  1.5&  --7.9&   1.6& --123  &  50  \\ 
\hline 
\end{tabular}          
\end{table*}
\end{landscape}
\end{center}


\begin{thebibliography}{}
\bibitem[Alcock, Allsman \& Axelrod 1996]{aaa96} Alcock, C., Allsman, R.,   
     Axelrod, T., Bennett, D. et al. 1996, AJ, 111, 1146 
\bibitem[Alcock, Alves \& Axelrod 2004]{aaa04} Alcock, C., Alves, D., 
     Axelrod, T., Becker, T. et al.  2004, AJ, 127, 334  
\bibitem[Altmann, Catelan \& Zoccali 2005]{acz05} Altmann, M., Catelan, M. \& 
     Zoccali, M. 2005, A \& A, 439, 5 
\bibitem[Battaglia et al. 2005]{bat05} Battaglia, G., Helmi, A., Morrison, H. 
     et al. 2005, MNRAS, 364, 433		
\bibitem[Beers et al. 1996]{b96} Beers, T.C., Wilhelm, R., Dionidis, S.P. \& 
     Mattson, C.J. 1996, ApJ Suppl., 103, 433     
\bibitem[Beers et al. 2000]{b00} Beers, T.C., Chiba, M., Yoshii, Y. et al.  
     2000, AJ, 119, 2866 
\bibitem[Belokurov et al. 2006]{bel06} Belokurov, V., Zucker, D., Evans, N.,   
        Gilmore, G. et al. 2006, ApJ, 642, L137  
\bibitem[Bessell, et al. 1998]{bcp98}Bessell, M., Castelli, F., Plez, B.
     1998, A\&A, 333, 231  
\bibitem[Borkova \& Marsakov]{bm03} Borkova, T., Marsakov, V. 2003, A\&A,
         398, 133 
\bibitem[Brown et al. 2003]{br03} Brown,W.R., Geller, M.J., Kenyon, S.J. 
     et al. 2003, AJ, 126, 1362     
\bibitem[Brown et al. 2004]{br04} Brown,W.R., Geller, M.J., Kenyon, S.J. 
     et al. 2004, AJ, 127, 1555     
\bibitem[Brown et al. 2005]{br05} Brown,W.R., Geller, M.J., Kenyon, S.J. 
     et al. 2005, AJ, 130, 1097 
\bibitem[Cardelli, Clayton \& Mathis 1989]{ccm89} Cardelli, J.A., Clayton, 
     G.C. \& Mathis, J.S. 1989, ApJ, 345, 245     
\bibitem[Carney(1999)]{car99}  
	Carney, B. W. 1999, in {\it The Third Stromlo Symposium: 
	The Galactic Halo}, eds. Gibson, B. K., Axelrod, T. S.,  Putman, M. E., ASP 
	Conf.\ Ser.\ Vol.\ 165, p.\ 230
\bibitem[Chiba \& Beers(2000)]{chi00}  
	Chiba, M.,  Beers, T. C. 2000, AJ, 119, 2843	
\bibitem[Chiba \& Mizutani 2004]{chi04}  
	Chiba, M. \&  Mizutani, A. 2004, P.A.S. Australia, 21, 237 	
\bibitem[Christlieb et al. 2005]{chr05} Christlieb, N., Beers, T. \& Thom, C.
     2005, A\&A, 431, 143      
\bibitem[Clement et al. 2001]{cmd01} Clement, C., Muzzin, A., Dufton, Q 
     et al. 2001 AJ, 122, 2587  
\bibitem[Clementini et al. 2003]{cle03} Clementini, G., Gratton, R., 
     Bragaglia, A., Carretta, E., Di Fabrizio, L. \& Maio, M. 2003, AJ, 125, 
     1309      
\bibitem[Clewley et al. 2002]{cle02} Clewley, L. et al. 2002, MNRAS, 337, 87
\bibitem[Clewley et al. 2004]{cle04} Clewley, L. et al. 2004, MNRAS, 352, 285
\bibitem[Clewley et al. 2005]{cle05} Clewley, L. et al. 2005, MNRAS, 362, 349
\bibitem[Clewley \& Kinman 2006]{cle06} Clewley, L. \& Kinman, T.D. 2006, 
     MNRAS, 371, L11  
\bibitem[Contreras et al. 2005]{con05} Conteras, R., Catelan, M., Smith, H.
        et al. 2005, ApJ, 623, 117
\bibitem[Cseresnjes 2003]{cse03} Cseresnjes, P. 2003, A \& A, 375, 909 
\bibitem[De Angeli et al. 2005]{dea05} De Angeli, F., Piotto, G., Cassisi, S., 
      Busso, G. et al.  2005, AJ, 130, 116 
\bibitem[Dehnen \& Binney(1998)]{db98} 
	Dehnen, W.,  Binney, J. J. 1998, MNRAS, 298, 387
\bibitem[Duffau et al. 2006]{duf06} Duffau,S., Zinn, R., Vivas, A.K.   
     et al., 2006, ApJ, 636, L97
\bibitem[Ferraro et al. 1997]{fer97} Ferraro, F.R., Carretta, E., Corsi, C.E., 
     Fusi Pecci, F., Cacciari, C., Buonanno, R., Paltrinieri, B. \& Hamilton, D.
     1997, A \& A, 320, 757 
\bibitem[Freedman et al. 2001]{fr01} Freedman, W.L. et al. 2001, ApJ, 553, 47 
\bibitem[GAIA, Concept and Technology Study Report 2004]{gaia04} GAIA, Concept 
     and Technology Study Report 2004, ESA-SCI(2004)4 
\bibitem[Gilmore, Wyse \& Norris(2002)]{gil02} 
	Gilmore, G., Wyse, R. F. G.,  Norris, J. E. 2002, ApJ, 574, L39	
\bibitem[Helmi et al. 1999]{hel99} Helmi, A., White, S.D., de Zeeuw, P.T. 
     \& Zhao, H. 1999, Nature, 402, 53 
\bibitem[Harding et al. 2001]{har01}  Harding, P., Morrison, H.L., Olszewski, E.W., 
     Arabadjis, J., Mateo, M., Dohm-Palmer, R.C., Freeman, K.C. \& 
     Norris, J.E. 2001, AJ, 122, 1397    
\bibitem[Hewitt \& Burbidge]{heb93} Hewitt, A., Burbidge, G., 1993, ApJS, 87,
        451
\bibitem[Ibata, Gilmore \& Irwin 1994]{iba94} Ibata, R., Gilmore, G. \& 
      Irwin, M. 1994, Nature, 370, 194     
\bibitem[Ibata et al. 2001]{iba01} Ibata, R., Lewis, G.F., Irwin, M., 
     Totten, E.J. \& Quinn, T. 2001, ApJ, 551, 294     
\bibitem[Ivezi\'c et al. 2000]{iv00} Ivezi\'c, Z., Goldston, J., Findlator, K. 
     et al. 2000, AJ, 120, 963 
\bibitem[Johnson \& Soderblom(1987)]{john87}  
	Johnson, D. R. H,,  Soderblom, D. R. 1987,  AJ, 93, 864	
\bibitem[Jones \& Walker 1988]{jw88} Jones, B.F. \& Walker, M.F. 1988, AJ, 95, 1755
\bibitem[Jones et al. 1996]{j96} Jones, R.V., Carney, B.W. \& Fulbright, J.P. 
     1996, PASP, 108, 877 	
\bibitem[Juric et al. 2006]{jur06} Juri\'c, M., Ivezi\'c, Z., Brooks, A. et al. 
     2006, ApJ, subm. (astro-ph/0510520) 
\bibitem[Kepley \& al. 2006]{kep06} Kepley, A. Morrison, H. Helmi, A.,       
         Kinman, T. et al. 2006, AJ,{\em submitted}. 
\bibitem[Kholopov et al. 1985]{kho85} Kholopov, P. et al. 1985 Gen. Cat. of
       Variable Stars (http://www.sai.msu.ru/gcvs/gcvs/) 
\bibitem[Kinman et al. 1965]{kin65} Kinman, T.D., Wirtanen, C.A., Janes, K.A.,
         1965, ApJS, 11, 223  
\bibitem[Kinman, Suntzeff \& Kraft 1994]{ksk94} Kinman, T.D., Suntzeff, N.B. \& 
     Kraft, R.P. 1994, AJ, 108, 1722
\bibitem[Kinman et al.(1996)]{kin96}  
	Kinman, T.D., Pier, J.R., Suntzeff, N.B., {\it et al.} 1996, AJ, 111, 
	1164 
\bibitem[Kinman et al. (2000)]{kin00}  
	Kinman, T. D., Castelli, F.,  Cacciari, C., Bragaglia,  A., Harmer, D.,
        Valdes, F. 2000, A\&A, 364, 102  
\bibitem[Kinman(2002)]{kin02a}  Kinman, T. D. 2002, IBVS 5311	
\bibitem[Kinman et al. (2003)]{kin03}  
	Kinman, T. D.,  Cacciari, C., Bragaglia,  A., Buzzoni, A., Spagna, A. 
	2003, in {\it Galactic,  Stellar Dynamics}, Proc.\ of JENAM 2002, 
	eds.\ C. M. Boily, P. Patsis, S. Portegies-Zwart, R. Spurzem,  
	C. Theis, EAS Pub. Ser.,  Vol.\ 10, p.\ 115	
\bibitem[Kinman et al.(2005)]{kin05}  
	Kinman, T.D.,  Bragaglia,  Cacciari, C., A., Buzzoni, A., Spagna, A. 
	2005, in {\it The Three-Dimensional Universe with GAIA}, ESA SP-576,  
	p.\ 175		
\bibitem[Kinman et al. 2007]{kin06} Kinman, T.D. et al. 2007, ({\em in preparation}) 
\bibitem[Lasker et al. 1995]{las05} Lasker, B.M., McLean, B.J., Jenkner, H., 
     Lattanzi, M.G. \& Spagna, A. 1995, in {\it Future Possibilities for
     Astrometry in Space}, eds. M.A.C. Perryman, F. van Leeuwen, T.D. Guyenne, 
     ESA SP-379, 137
\bibitem[Law, Johnson \& Majewski 2005]{ljm05} Law, D.R., Johnson, K.V. \& 
     Majewski, S. 2005, ApJ, 619, 807      
\bibitem[Layden A. 1996]{l96} Layden, A. 1996, 
                   in {\it The Formation of the Galactic Halo.. Inside and
	out }, eds. Morrison, H. \& Sarajedini, A.,  ASP 
	Conf.\ Ser.\ Vol.\ 92, p.\ 141
\bibitem[Lee J-W. \& Carney, B.W. 1999]{lc99} Lee, J-W., Carney, B.W. 1999,
           AJ, 118, 1373  
\bibitem[Lupton et al. 2005]{lup05} Lupton, R., Juric, M., Ivezic, Z., 
           Brooks, A. et al. 2005, BAAS, 37, 1384  
\bibitem[MacConnell et al. 1993]{mac93} MacConnell, D.J., Stephenson, C.B. \&  
     Pesch, P. 1993, ApJS, 86, 453      
\bibitem[Maintz \& de Boer 2005]{mdb05} Maintz, G. \& de Boer, K.S. 2005, 
     A \& A, 442, 229 
\bibitem[Majewski 1992]{maj92}  
	Majewski, S.R. 1992, ApJS, 78, 87 	
\bibitem[Majewski 2005]{maj05} Majewski, S.R. 2005, ({\em priv. comm.}) 
\bibitem[Majewski, Munn \& Hawley(1996)]{maj96}  
	Majewski, S.,  Munn, J. A., Hawley, S. L. 1996, ApJ, 459, L73	
\bibitem[Majewski et al. 2003]{maj03}  Majewski, S.R., Skrutskie, M.F., 
     Weinberg, M.D. \& Ostheimer, J.C. 2003, ApJ, 599, 1082       
\bibitem[Martin \& Morrison(1998)]{mar98}  
	Martin, J. C., Morrison, H. L. 1998, AJ, 116, 1724	
%\bibitem[Martin et al. 2004]{mar04}  Martin, N.F., Ibata, R.A., Bellazzini, 
%     M. et al.  2004, MNRAS, 348, 12
%\bibitem[Mart\'{\i}nez-Delgardo et al. 2005]{md05} Mart\'inez-Delgardo, D., 
%     Butler, D.J., Rix, H-W. et al. 2005, ApJ, 633, 205	
\bibitem[Mart\'{\i}nez-Delgardo et al. 2006]{md06} Mart\'inez-Delgardo, D., 
       Pe\~{n}arrubia, J., Juric, M. et al. 2006,  astro-ph/0609104  
\bibitem[McLean et al. 2000]{mcl00} McLean, B.J., Greene, G.R., Lattanzi, M.G. 
     \& Pirenne, B. 2000, in {\it ADASS IX}, eds. N. Manset, C. Veillet, 
     D. Crabtree, ASP Conf. Ser. Vol. 216, 145       
\bibitem[Meza et al. 2005]{mez05} Meza, A., Navarro, J.F., Abadi, M.G. \& 
     Steinmetz, M. 2005, MNRAS, 359, 93      
\bibitem[Monaco et al. 2003]{mon03} Monaco, L., Bellazzini, M., Ferraro, F.R. 
     \& Pancino, E. 2003, ApJ, 597, L25     
\bibitem[Morrison et al. 2000]{mor00} Morrison, H.L., Mateo, M., Olszewski, 
     E.W., Harding, P., Dohm-Palmer, R.C., Freeman, K.C., Norris, J.E. \& 
     Morita, M. 2000, AJ, 119, 2254     
\bibitem[Nemec et al. 1994]{nem04} Nemec, J.M., Linnell-Nemec, A.F. \& 
     Lutz, T.E. 1994, AJ, 108, 222     
\bibitem[Newberg et al. 2002]{new02} Newberg, H.J., Yanny, B., Rockosi, C. 
     et al. 2002, ApJ, 569, 245      
\bibitem[Paltrinieri et al. 1998]{pal98} Paltrinieri, B., Ferraro, F.R., Fusi 
     Pecci, F. \& Carretta, E. 1998, MNRAS, 293, 434       
\bibitem[Preston, Shectman \& Beers(1991)]{pre91}  
	Preston, G. W., Shectman, S. A., Beers, T. C. 1991, ApJ, 375, 121	
\bibitem[Rosenberg et al. 1999]{ros99}Rosenberg, A., Saviane, I., Piotto, G.,
        Aparicio, A. 1999, AJ, 118, 2306
\bibitem[Sandage 2006]{san06} Sandage, A. 2006, AJ, 131, 1750	
\bibitem[Sanduleak 1988]{s88} Sanduleak, N. 1988, ApJS, 66, 309 	
\bibitem[Schlegel, Finkbeiner \& Davies(1998)]{schle98}  
	Schlegel, D. J.,  Finkbeiner, D. P., Davis, M. 1998, ApJ, 500, 525
\bibitem[Shapiro \& Wilk 1965]{sw65} Shapiro, S.S. \& Wilk, M.B. 1965, 
     Biometrika, 52, 591  	
\bibitem[Sirko et al. 2004a]{sir04a} Sirko, E., Goodman, J., Knapp, G.R. et al. 
     2004a, AJ, 127, 899 
\bibitem[Sirko et al. 2004b]{sir04b} Sirko, E., Goodman, J., Knapp, G.R. et al. 
     2004b, AJ, 127, 914 
\bibitem[Smith et al. 2006]{smi06} Smith M., Ruchti, G., Helmi, A., Wyse, R.
         et al. 2006, astro-ph/0611671 
\bibitem[Soszy\'{n}ski et al. 2003]{sus03} Soszy\'{n}ski, I., Udalski, A.
     Syma\'{n}ski M. et al. 2003, Acta Astron., 53, 93
\bibitem[Spagna et al.(1996)]{spa96} 
	Spagna, A., Lattanzi, M. G., Lasker, B. M., McLean, B. J., Massone, G., 
	 Lanteri, L. 1996, A\&A, 311, 758     
\bibitem[Stetson 1991]{ste91} Stetson, P. 1991, AJ, 102, 589 
\bibitem[Takase \& Miyauchi-Isobe 1993]{tak93} Takase, B., Miyauchi-Isobe, N.
         1993 Publ. Natl. Astron. Obs. Japan, 3, 169  
\bibitem[Tammann, Sandage \& Reindl 2003]{tsr03} Tammann, G.A., Sandage, A. 
     \& Reindl, B. 2003, A\&A, 404, 423	
\bibitem[Thom et al. 2005]{tho05} Thom, C., Flynn, C., Bessell, M.S., et al. 
     2005, MNRAS, 360, 354     
\bibitem[Totten \& Irwin 1998]{ti98} Totten, E.J. \& Irwin, M.J. 1998, MNRAS, 
     294, 1      
\bibitem[Totten, Irwin \& Whitelock 2000]{tiw00} Totten, E.J., Irwin, M.J. \& 
     Whitelock, P.A. 2000, MNRAS, 314, 630      
\bibitem[Valenti et al. 2004]{val04} Valenti, E., Ferraro, F.R., Perina, S. 
     \& Origlia, L. 2004, A\&A, 419, 139  
\bibitem[Vallenari et al. 2006]{vall06} Vallenari, A., Pasetto, S., Bertelli, 
     G., Chiosi, C., Spagna, A. \& Lattanzi, M. 2006, A\&A, 451, 125      
\bibitem[van den Bergh, 1993]{vdb93} van den Bergh, S. 1993, AJ, 105, 971
\bibitem[VandenBergh et al. 2000]{vdb00}VandenBergh, D., Swenson, F., Rogers, 
      F. et al.  2000,  ApJ, 532, 430 
\bibitem[V\'{e}ron-Cetty  \& V\'{e}ron, 2001]{vcv01} V\'{e}ron-Cetty, M.,
   V\'{e}ron, P.,  2001 CDS VizieR (VII/224) 
\bibitem[Vivas et al. 2001]{viv01} Vivas, A.K., Zinn, R., Andrews, P. et al. 
     2001, ApJ, 554, L33      
\bibitem[Vivas \& Zinn 2006]{viv06} Vivas, A.K., Zinn, R., 2006, AJ, 132, 714 
\bibitem[Wesselink, Th. 1987]{wes87} Wesselink, Th., 1987, PhD. thesis,
       Catholic University of Nijmegen, The Netherlands 
\bibitem[Wilhelm et al. 1996]{wil96} Wilhelm, R., Beers, T., Kriessler, J.
     et al. 1996,  in {\it The Formation of the Galactic Halo.. Inside and
	out }, eds. Morrison, H. \& Sarajedini, A.,  ASP 
	Conf.\ Ser.\ Vol.\ 92, p.\ 171
\bibitem[Yanny et al. 2000]{ya00} Yanny, B., Newberg, H.J., Kent, S. et al. 
     2000, ApJ, 540, 825 
%\bibitem[Yanny et al. 2003]{ya03} Yanny, B., Newberg, H.J., Grebel, E.K. et al. 
%     2003, ApJ, 588, 824 
\bibitem[Zacharias et al. 2004]{zac04} Zacharias, N., Monet, D., Levine, S., et
  al. 2004, BAAS, 36, 1418 
\bibitem[Zinn 1993]{z93} Zinn, R.  1993, in {\it The Globular Cluster $-$      
   Galaxy Connection }, eds. Smith, G \& Brodie, J., ASP 
	Conf.\ Ser.\ Vol.\ 48, p.\ 38
\end{thebibliography}
\end{document}